\documentclass[journal]{IEEEtran}

\usepackage{amsmath,amssymb,amsfonts,amsthm,mathrsfs}
\usepackage{algorithm,algorithmic}
\usepackage{graphicx}
\usepackage{textcomp}
\usepackage{color,xcolor}
\usepackage{bm}
\usepackage{multirow}
\usepackage[normalem]{ulem}
\usepackage{tabularx}
\usepackage{soul}
\usepackage{ulem}
\usepackage{verbatim}
\usepackage{subcaption}
\usepackage{enumitem}
\usepackage{hyperref}
\usepackage[acronym]{glossaries}
\usepackage[linewidth=0.8pt]{mdframed}
\usepackage{cite}
\hypersetup{
	colorlinks=true,  
	linkcolor=red,    
	citecolor=blue,   
	 urlcolor=black   
}

\usepackage{geometry}
\theoremstyle{theorem,lemma,remark,proposition}

\newtheorem{remark}{Remark}

\ifCLASSINFOpdf
\else
\fi

\begin{document}

\title{
Harnessing Rydberg Atomic Receivers: From Quantum Physics to Wireless Communications
}

\author{
	Yuanbin~Chen,~Xufeng~Guo,~Chau~Yuen,~\IEEEmembership{Fellow,~IEEE},~Yufei~Zhao,~Yong~Liang~Guan,~\IEEEmembership{Senior~Member,~IEEE},~Chong~Meng~Samson~See,~M\'{e}rouane~Debbah,~\IEEEmembership{Fellow,~IEEE},\protect\\and~Lajos~Hanzo,~\IEEEmembership{Life~Fellow,~IEEE}

\vspace{-1em}

	\thanks{
		This work was supported in part by Singapore Ministry of Education (MOE) Academic Research Fund (AcRF) Tier 1 Thematic Grant RT12/23 023780-00001. The financial support of the following Engineering and Physical
		Sciences Research Council (EPSRC) projects is gratefully acknowledged:
		Platform for Driving Ultimate Connectivity (TITAN) (EP/X04047X/1;
		EP/Y037243/1); Robust and Reliable Quantum Computing (RoaRQ,
		EP/W032635/1); PerCom (EP/X012301/1); EP/X01228X/1; India-UK Intelligent
		Spectrum Innovation ICON UKRI-1859. \textit{(Corresponding authors: Lajos Hanzo; Chau Yuen.)}

		Yuanbin Chen, Xufeng Guo, Chau Yuen, Yufei Zhao, and Yong Liang Guan are with the School of Electrical and Electronics Engineering, Nanyang Technological University, Singapore 639798 (emails: yuanbin.chen@ntu.edu.sg; n2308905j@e.ntu.edu.sg; chau.yuen@ntu.edu.sg; yufei.zhao@ntu.edu.sg; eylguan@ntu.edu.sg).
		
		Chong Meng Samson See is with DSO National Laboratories, Singapore 118225 (e-mail: schongme@dso.org.sg).
		
		M\'{e}rouane Debbah is with the Research Institute for Digital Future, Khalifa University, 127788 Abu Dhabi, UAE (e-mail: merouane.debbah@ku.ac.ae). 
		
		Lajos Hanzo is with School of Electronics and
		Computer Science, University of Southampton, SO17 1BJ Southampton, U.K. (e-mail: lh@ecs.soton.ac.uk).		
	}
	
}

\maketitle

\begin{abstract}
The intrinsic integration of Rydberg atomic receivers into wireless communication systems is proposed, by harnessing the principles of quantum physics in wireless communications. More particularly, we conceive a pair of Rydberg atomic receivers, one incorporates a local oscillator (LO), referred to as an LO-dressed receiver, while the other operates without an LO and is termed an LO-free receiver. The appropriate wireless model is developed for each configuration, elaborating on the receiver's responses to the radio frequency (RF) signal, on the potential noise sources, and on the signal-to-noise ratio (SNR) performance. The developed wireless model conforms to the classical RF framework, facilitating compatibility with established signal processing methodologies.
Next, we investigate the associated distortion effects that might occur, specifically identifying the conditions under which distortion arises and demonstrating the boundaries of linear dynamic ranges. This provides critical insights into its practical implementations in wireless systems. Finally, extensive simulation results are provided for characterizing the performance of wireless systems, harnessing this pair of Rydberg atomic receivers. Our results demonstrate that LO-dressed systems achieve a significant SNR gain of approximately $40\sim50$~dB over conventional RF receivers in the standard quantum limit regime. This SNR head-room translates into reduced symbol error rates, enabling efficient and reliable transmission with higher-order constellations.


\end{abstract}


%
\IEEEpeerreviewmaketitle

\section{Introduction}
Quantum mechanics, emerging in the early 20th century, constitute one of the foundational pillars of modern physics, reshaping our understanding of the nanoscale world. By harnessing the alluring principles of quantum theory, quantum technology has revolutionized our ability to manipulate and control subatomic particles with unprecedented precision, catalyzing advances in communications, computing, and especially precision measurement~\cite{QS-101,QS-104,QS-BG1,QS-22}.
One of most pivotal branches of quantum technology is \textit{quantum sensing} capable of measuring physical quantities with extraordinary sensitivity and accuracy~\cite{QS-30,QS-31,QS-BG2}.  Rapidly establishing itself as an evolving research frontier, quantum sensing might rely on spin qubits, trapped ions, and flux qubits, positioning itself as a game changer in unlocking new possibilities for applied physics and wireless communications. As quantum sensing continues to evolve, it promises to redefine the boundaries of measurement precision, paving the way for breakthroughs that will significantly enhance the performance of the next-generation wireless communications.


\subsection{Fundamentals of Rydberg Atoms}
Rydberg atoms are characterized by their excitation to certain energy states with large principal atomic numbers,
having at least one electron in a highly excited state. This excitation results in having the outer electron far from the ionic core of the Rydberg atom, with the associated dipole moment roughly scaling with its principal atomic number. Having a high electric dipole moment enables Rydberg atoms to engage in strong coupling with even extremely weak radio frequency (RF) fields, making them high-sensitivity sensors for electric fields, particularly for detecting time-varying signals. Specifically, in response to an external RF electric field, a Rydberg atom becomes coupled with this RF field through its high electric dipole moment, and hence the state of the Rydberg atom changes. This event triggers atomic energy level transitions and causes fluctuations in population distribution among these states. These perturbations are precisely monitored by the so-called electromagnetically induced transparency (EIT), which is an all-optical readout technique that probes the Rydberg atomic state~\cite{QS-22}.
Briefly, the atomic energy levels variations caused by the interaction with the RF field results in EIT spectrum changes, allowing for the accurate extraction of the information conveyed by the electric field. This intricate interplay between the intrinsic properties of Rydberg atoms and EIT-based detection constitutes the fundamental principle that enables Rydberg atoms to function as high-sensitivity quantum sensors.

\subsection{State-of-the-Art in Rydberg Atomic Receivers}

For over four decades, Rydberg atoms have captivated researchers as promising candidates for electric field sensing applications~\cite{QS-22-2, QS-22-3, QS-22-4}. The realm of Rydberg atom-based electrometry, dedicated to the precise metrology of electric fields, witnessed a period of rapid development between 2010-2014, underscored by several important studies~\cite{QS-22-5, QS-12, QS-10-TAP}. Specifically, in~\cite{QS-22-5}, the idea of employing Rydberg atoms for electric field measurements was proposed, paving the way for establishing the linkage between the optical response characteristics and the International System of Units (SI)~\cite{QS-22-9}. This direct linkage eliminated the need for additional calibration and hence facilitated absolute measurements. This theoretical innovation was experimentally verified in \cite{QS-12}, while the authors of~\cite{QS-10-TAP} further demonstrated the associated self-calibrated traceability over a broad frequency range. Based on these inspirational achievements, numerous academic and industrial research initiatives have launched Rydberg atomic sensor programs across the globe, 
highlighting the benefits of their sensitivity and versatility, thus positioning them at the forefront of next-generation quantum sensing research.

At the heart of Rydberg atomic receivers lies (i) the ability to diagnose the phase and frequency of weak RF electric fields, and (ii) the enhancement of detection sensitivity. Based on their different responses to RF fields, Rydberg atomic receivers can be categorized into \textit{local oscillator (LO)-free}~\cite{QS-10-TAP,QS-11,QS-12,QS-15,QS-24,QS-4} and \textit{LO-dressed} contexts~\cite{QS-9,QS-16,QS-26,QS-27,QS-28}. To be specific, LO-free Rydberg atomic receivers, also referred to as the electrometry, measure the amplitude of the RF field by exploring the so-called Autler-Townes (AT) splitting effect~\cite{QS-10-TAP,QS-11,QS-12,QS-15,QS-24,QS-4}.
By measuring the interval of the two split peaks of the EIT spectrum, the Rabi frequency of the RF field can be inferred, thereby allowing for the identification of the RF electric field strength with a high sensitivity approaching the standard quantum limit (SQL). Quantitatively, this is on the order of ${\mu\text{V/cm/}}\sqrt {{\text{Hz}}} $~\cite{QS-3,QS-19}. However, this method is limited to probing the amplitude and polarization of the electric field through optical readout, but fails to measure the phase of the RF signal~\cite{QS-12,QS-24}.

By contrast, the LO-dressed Rydberg atomic receiver incorporates an additional RF field that serves as an LO, which is on-resonance with the Rydberg transition. This setup relies on both the EIT and AT effects in Rydberg atoms in order to demodulate a secondary, co-polarized RF field, hence facilitating precise phase measurements by treating the atomic system as a Rydberg atomic mixer~\cite{QS-9,QS-16,QS-26,QS-27,QS-28}. These are also referred to as atomic superheterodyne receivers~\cite{QS-16, QS-28}. The resultant LO-dressed entities achieve an unparalleled sensitivity, capable of detecting RF electric fields on the order of $\text{nV/cm}/\sqrt{\text{Hz}}$.
Consequently, the Rydberg atomic receiver behaves like an integrated compact high-sensitivity antenna, adept at the detection and reception of both amplitude-modulated (AM)~\cite{QS-10-APM,QS-18,QS-22-15}, frequency-modulated (FM)~\cite{QS-10-APM,QS-22-15}, and phase-modulated (PM) signals~\cite{QS-6,QS-22-19}.

\subsection{The Advantages}

Compared to conventional RF receivers, which rely on traditional antennas and complex electronic front-end components, Rydberg atomic receivers exhibit several remarkable advantages. Firstly, traditional RF receivers utilize antennas to absorb the impinging electromagnetic energy, converting free-space modes into guided currents that are subsequently amplified, filtered, and rectified by front-end circuits before being processed in the analog or digital domains. By contrast, Rydberg atomic receivers do not require any net absorption of the incoming RF field. Instead, the incident RF field alters the energy levels of Rydberg atoms through coherent interactions, which is then imparted to an optical field as amplitude or phase modulations and detected spectroscopically~\cite{QS-3}. This eliminates the need for complex and bulky front-end electronic circuits.
Secondly, Rydberg atomic receivers are inherently wavelength-independent~\cite{QS-22-20}, unlike traditional antennas that must be comparable in size to the wavelength of the RF signal. This feature enables the fabrication of highly efficient electrically small antennas using Rydberg atoms. Thirdly, while conventional high-frequency antennas are inherently directional, Rydberg atomic receivers maintain an omni-directional profile across all wavelengths~\cite{QS-32-A}.
Last but not least, Rydberg atomic receivers embrace a broad detection frequency range. The extensive energy level structure of Rydberg atoms allows for their coupling with electromagnetic waves spanning from direct current to Terahertz (THz) frequencies by merely adjusting these energy levels. This tunability facilitates the sensing of various RF bands without necessitating reconfiguration of the hardware platform, which is a significant challenge for traditional RF receivers.

\subsection{Motivation and Our Contributions}

Despite the significant advances concerning Rydberg atomic receivers in the context of quantum sensing, their application in wireless communications is still in its infancy~\cite{QS-101,QS-1,xinyi_R1,terry-mag,terry-trans}. Our goal in this treatise is to bridge the gap between quantum physics and wireless communications by developing bespoke mathematical models for both LO-free and LO-dressed Rydberg atomic receivers, in order to unveil their great potentials and provide a solid foundation for their potent applications in wireless systems.
To this end, this very first study aims for filling the knowledge gap in the-state-of-the-art through the following contributions.
\begin{itemize}
\item Starting from quantum physics, we develop appropriate wireless models tailored for both LO-free and LO-dressed Rydberg atomic receivers, specifically focusing on the receivers' responses to the RF signal. Given their different operating mechanisms, the potential noise sources inherent in each receiver type are explicitly examined. The signal model developed conforms to the classical RF framework, facilitating compatibility with established signal processing methodologies.

\item We investigate the distortion effects that might occur when deploying both LO-free and in LO-dressed Rydberg atomic receivers. Specifically, for LO-free systems, the distortion is induced by the ambiguous observation in AT splitting owing to the weak Rabi frequency of the RF field. We introduce a quantitative threshold based on the ratio of the RF Rabi frequency to the EIT linewidth; below this ratio the optical readout becomes ambiguous. As regards to LO-dressed systems, we analytically identify the upper and lower bounds of the linear dynamic range over which the probe transmission remains linear in terms of the Rabi frequency.
Furthermore, we provide their respective signal-to-noise ratios (SNR) expressions that explicitly characterize these distortion behaviors, providing an intuitive understanding of their achievable performance in practical implementations.

\item By leveraging the QuTiP toolkit~\cite{QuTip}, we design quantum systems for supporting both LO-free and LO-dressed Rydberg atomic receivers. Specifically, three key metrics are adopted for performance evaluations, namely the SNR, mutual information, and symbol error rate (SER). 
Our numerical results demonstrate that, for a given thermal-noise floor, the LO-dressed system achieves the best SNR performance among all configurations considered. When the quantum system is pushed to the SQL regime, the LO-dressed Rydberg atomic receiver attains an SNR gain of approximately $40\sim50~\text{dB}$ over the conventional RF receiver. Additionally, the LO-dressed front-end minimizes the received power necessary to achieve a specific SER, with the required power decreasing further under SQL constraints in comparison to its LO-free and classical RF counterparts.

\end{itemize}

This paper is organized as follows. In Sec.~\ref{section_preliminary}, we present the preliminaries, detailing the fundamental operating principles of Rydberg atomic receivers to establish the necessary groundwork for the subsequent discussions. In Sec.~\ref{section_modeling_from_phy_to_wireless}, a pair of models tailored specifically for LO-free and LO-dressed Rydberg atomic receivers are developed, designed for integration into wireless communication systems. In Sec.~\ref{section_distortion_effect}, we investigate the distortion effects that may occur when these receivers are employed in practical wireless systems, elucidating the distinct challenges associated with each receiver type. Sec.~\ref{section_numerical_results} provides numerical results, demonstrating some interesting findings and illustrating the performance of these receivers in various wireless scenarios. Finally, Sec.~\ref{section_conclusion} concludes this paper.
%


\section{Preliminaries}\label{section_preliminary}

To provide an intuitive grasp of the Rydberg atomic receiver concept, this section commences by discussing a key physical phenomenon, namely EIT, which drives its operating principle and facilitates an accurate readout for RF field measurements.

\subsection{Electromagnetically Induced Transparency}

\begin{figure*}[t]
	\centering
	\includegraphics[width=0.95\textwidth]{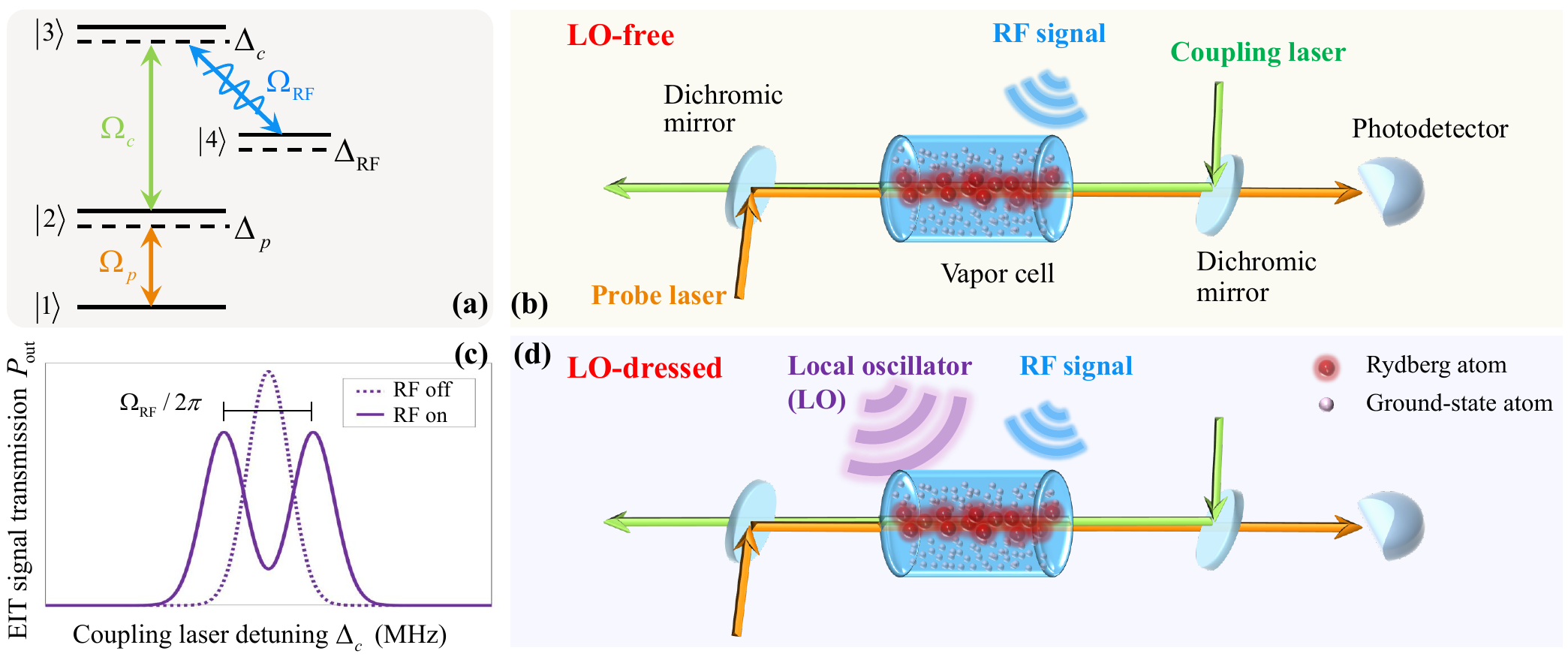}
	\caption{Illustration of the Rydberg atomic receivers and measurement principles. (a) Energy level diagram. (b) EIT and AT-splitting based measurement. (c) LO-free Rydberg atomic receiver. (d) LO-dressed Rydberg atomic receiver.} \label{system_model}
	\vspace{-1em}
\end{figure*}

Fig.~\ref{system_model} illustrates the excitation of atoms to Rydberg states within a vapor cell. The vapor cell is populated with Rydberg atoms such as Rubidium (Rb) or Cesium (Cs), which serve as the receiver medium. Within this setup, two counter-propagating laser, namely the probe laser and the coupling laser, are harnessed for manipulating the atomic states. The probe laser facilitates the coupling of the atomic ground state $\left| 1 \right\rangle $ to a low-lying excited state $\left| 2 \right\rangle $ (also referred to as the first excited state), while the coupling laser drives the transition from $\left| 2 \right\rangle $ to a highly excited Rydberg state $\left| 3 \right\rangle $. When both lasers are precisely resonant, the atoms are coherently pumped into a superposition of $\left| 1 \right\rangle $ and $\left| 3 \right\rangle $, rendering this state transparent to the probe laser - a phenomenon known as EIT~\cite{QS-16-23}. This reduction in probe absorption under resonant conditions is the hallmark of EIT. By scanning the coupling laser and monitoring the resultant probe absorption, the EIT signal effectively spectroscopically probes the energy of the Rydberg state $\left| 3 \right\rangle $, as the probe absorption decreases when the coupling laser is aligned with the transition energy.

Upon excitation to the Rydberg state $\left| 3 \right\rangle $, the atoms become highly sensitive to RF fields that are resonant or nearly resonant with transitions to adjacent Rydberg states, such as $\left| 4 \right\rangle $. When an RF field induces a resonant coupling between $\left| 3 \right\rangle $ and $\left| 4 \right\rangle $, the resultant AT effect manifests itself as a splitting of the energy levels, which in turn results in the bifurcation of the EIT signal peak, as depicted in Fig.~\ref{system_model}(c). The frequency difference between these two AT-split resonances, denoted by $\Delta \nu$, directly corresponds to the energy difference introduced by the RF field, which is given by
\begin{equation}\label{AT-interval}
\Delta \nu  = \left\{ \begin{gathered}
	\frac{{{\Omega _{{\text{RF}}}}}}{{2\pi }},{\text{   scanning the coupling laser,}} \hfill \\
	\frac{{{\Omega _{{\text{RF}}}}}}{{2\pi }}\frac{{{f_p}}}{{{f_c}}},{\text{   scanning the probe laser,}} \hfill \\ 
\end{gathered}  \right.
\end{equation}
where $\Omega_{\text{RF}}$ denotes the Rabi frequency of the RF signal, $f_p$ and $f_c$ represent the frequency of the probe and coupling laser, respectively. Scanning the coupling laser is often preferred, since this way provides a direct measurement of the energy of the Rydberg state $\left| 3 \right\rangle $ without any underlying absorption background.

\subsection{Probe Laser Transmission}

Given the above foundational understanding of the EIT phenomenon, we proceed to the mathematical portrayal of the EIT signal to unveil its defining characteristics.
Under the adiabatic approximation of the probe laser transmission, which is in essence given by the intensity ${P_{\text{out}}}\left( t \right)$ of the probe laser measured on the photodetector (PD), it can be characterized through the imaginary part of the susceptibility as~\cite{QS-16-23}
\begin{equation}\label{P_out}
{P_{\text{out}}}\left( t \right) = {P_{\text{in}}}\exp \left\{ { - k_pL\Im \left( {\chi \left( t \right)} \right)} \right\},
\end{equation}
where ${P_{\text{in}}}$ denotes the intensity of the incident probe laser associated with its Rabi frequency $\Omega_p$
\begin{equation}\label{P_in}
P_{\text{in}} = \frac{\pi }{{2{Z_0}}}{\left( {\frac{{d {\Omega _p}\hbar }}{{2\sqrt {{\wp _{12}}} }}} \right)^2}.
\end{equation}
In (\ref{P_out}) and (\ref{P_in}), $L$ is the length of the vapor cell, $k_p = \frac{{2\pi }}{{{\lambda _p}}}$ is the wavenumber (wavevector) of the probe laser having a wavelength of $\lambda_p$, $\hbar$ denotes the reduced Planck's constant, $Z_0$ represents the impedance in the free space, $d$ is the $1/e^2$ diameter of the probe laser, ${\wp _{12}} $ characterizes the dipole moment associated with the  $\left| 1 \right\rangle  - \left| 2 \right\rangle $ transition, and  $\chi \left( t \right)$ indicates the susceptibility, which is given by
\begin{equation}\label{chi}
\chi \left( t \right) = C_0 {\rho _{21}}\left( t \right),
\end{equation}
where $ C_0 = \frac{{ - 2{N_0}{\wp _{12}}}} {{{\epsilon_0}\hbar {\Omega _p}}}$. Furthermore, ${\rho _{21}}\left( t \right)$ denotes the instantaneous steady-state density matrix component associated with the  $\left| 1 \right\rangle  - \left| 2 \right\rangle $ transition, $\epsilon_0$ is the permittivity in vacuum, and $N_0$ represents the total density of atoms in the cell. 

\subsection{Master Equation}
Taking into account the associated spontaneous emission, the dynamics of the atomic system are characterized by the master equation for the four-level density matrix $\bm{\rho}$~\cite{QS-15}
\begin{equation}\label{master_equation}
\dot {\bm{\rho }} = \frac{{\partial \bm{\rho }}}{{\partial t}} =  - \frac{\jmath }{\hbar }\left[ {{\mathbf{H}},\bm{\rho }} \right] + \mathcal{L}.
\end{equation}
In the master equation (\ref{master_equation}), $\mathbf{H}$ represents the Hamiltonian of the atomic system of interest, which is given by
\begin{align}\label{Hamiltonian}
	\resizebox{\hsize}{!}{$
{\mathbf{H}} = \frac{\hbar }{2}\left[ {\begin{array}{*{20}{c}}
		0&{{\Omega _p}}&0&0 \\ 
		{{\Omega _p}}&{ - 2{\Delta _p}}&{{\Omega _c}}&0 \\ 
		0&{{\Omega _c}}&{ - 2\left( {{\Delta _p} + {\Delta _c}} \right)}&{\Omega} \\ 
		0&0&{ \Omega  }&{ - 2\left( {{\Delta _p} + {\Delta _c} + {\Delta _{\text{RF}}}} \right)} 
\end{array}} \right], 
$}
\end{align}
where $\Omega_{p,c}$ and $\Delta_{p,c} $ indicate the Rabi frequencies and detunings associated with the probe laser and coupling laser, respectively, as well as ${\Delta _{\text{RF}}}$ represents the detuning of the RF signal. Note that the Rabi frequency $\Omega$ in (\ref{Hamiltonian}) differs for LO-free and LO-dressed Rydberg atomic receivers, which will be elaborated upon later. 
Still referring to (\ref{master_equation}), $\mathcal{L}$ denotes the Lindblad operator that encompasses the decay processes, which is given by
\begin{align}\label{Lindblad}
	\resizebox{\hsize}{!}{$
\mathcal{L} = \left[ {\begin{array}{*{20}{c}}
		{{\gamma _2}{\rho _{22}}}&{ - {\gamma _{12}}{\rho _{12}}}&{ - {\gamma _{13}}{\rho _{13}}}&{ - {\gamma _{14}}{\rho _{14}}} \\ 
		{ - {\gamma _{21}}{\rho _{21}}}&{{\gamma _3}{\rho _{33}} - {\gamma _2}{\rho _{22}}}&{ - {\gamma _{23}}{\rho _{23}}}&{ - {\gamma _{24}}{\rho _{24}}} \\ 
		{ - {\gamma _{31}}{\rho _{31}}}&{ - {\gamma _{32}}{\rho _{32}}}&{{\gamma _4}{\rho _{44}} - {\gamma _3}{\rho _{33}}}&{ - {\gamma _{34}}{\rho _{34}}} \\ 
		{ - {\gamma _{41}}{\rho _{41}}}&{ - {\gamma _{42}}{\rho _{42}}}&{ - {\gamma _{42}}{\rho _{42}}}&{ - {\gamma _4}{\rho _{44}}} 
\end{array}} \right].
$}
\end{align}
In (\ref{Lindblad}), we have ${\gamma _{ij}} = \left( { {\gamma _i} + {\gamma _j} } \right)/2$, where  ${\gamma _i}\left( {i = 1,2,3,4} \right)$ is the transition decay rate. The master-equation model treats each atom independently and atom-atom interactions can be neglected because the probe laser beam operates in the weak-excitation limit~\cite{QS-15,QS-16}.

The master equation in (\ref{master_equation}) is intractable, hence it does not have an analytical solution. Therefore, there is no closed-form expression for ${\rho _{21}}\left( t \right)$. We herein focus on its steady-state solution, achievable by constructing a matrix with the system of equations for $\dot {\bm{\rho}} = 0$. Specifically, a steady-state solution for $\rho_{21}$ can be given by
\begin{equation}\label{rho_21_LO_free}
{\rho _{21}} = \frac{{ - \jmath \left( {{\Omega _p}/2} \right)}}{{{\gamma _{21}} - \jmath {\Delta _p} - \frac{{{{\left( {{\Omega _c}/2} \right)}^2}}}{{{\gamma _{31}} - \jmath \left( {{\Delta _p} + {\Delta _c}} \right) - \frac{{{{\left( {{\Omega _{{\text{RF}}}}/2} \right)}^2}}}{{{\gamma _{41}} - \jmath \left( {{\Delta _p} + {\Delta _c} + {\Delta _{{\text{RF}}}}} \right)}}}}}}.
\end{equation}
See Appendix~\ref{appendix_rho} for detailed derivation. 
In a room-temperature vapor, the atom velocity $v_z$ follows a one-dimensional Maxwell-Boltzmann distribution~\cite{book_Maxwell-Boltzmann_distribution}
\begin{equation}
g\left( {{v_z}} \right) = \frac{1}{{\sqrt {2\pi } {\sigma _v}}}{e^{ - v_z^2/\left( {2\sigma _v^2} \right)}}.
\end{equation}
The corresponding standard deviation is $ {\sigma _v} = \sqrt {\frac{{{k_{\text{B}}}T}}{{{m_{{\text{Rb}}}}}}}  $, where $k_\text{B}$ is the Boltzmann constant, $T$ for the temperature, and $m_{\text{Rb}}$ for the mass of the atom. For the case where the probe and coupling laser are counter-propagating, the single-atom detuning encounters a Doppler shift
\begin{align}
{\Delta _p}\left( {{v_z}} \right) &= {\Delta _p} - \frac{{2\pi }}{{{\lambda _p}}}{v_z}, \nonumber\\
{\Delta _c}\left( {{v_z}} \right) &= {\Delta _c} + \frac{{2\pi }}{{{\lambda _c}}}{v_z},
\end{align}
with the sign set by the propagation direction. Replacing the detunings in (\ref{rho_21_LO_free}) yields a velocity-dependent coherence ${\rho_{21}}\left( {v_z} \right)$, which can be further Doppler averaged in the usual way~\cite{QS-13,xinyi_R2}
\begin{equation}\label{rho_21_LO_free_Doppler}
{\bar \rho _{21}}\left( {{\Delta _p}} \right) = \int_{ - \infty }^{ + \infty } {{\rho _{21}}\left( {{v_z}} \right)g\left( {{v_z}} \right)d{v_z}} .
\end{equation}

However, slight adjustments are required depending on the specifics of the Rydberg atomic regimes - regardless of LO-free or LO-dressed. Hence, the technique of measuring the signal's amplitude and phase using $\rho_{21} \left( t \right) $ also varies accordingly. The distinct Hamiltonians in both cases result in different forms of the density matrix component ${\bar \rho _{21}}\left( {{\Delta _p}} \right)$.
Thus, in the following sections, we present how $\rho_{21} \left( t \right) $ can be effectively harnessed for each scenario.

\section{Modeling From Quantum Physics To Wireless Systems}\label{section_modeling_from_phy_to_wireless}

This section presents the transitory solution of the Rydberg atomic receiver from its physical fundamentals to its application in a wireless system, relying on both LO-free and LO-dressed solutions. We commence by considering the transmitter (Tx) as a point source, i.e., by constructing a single-input and single-output (SISO) system. Then, we will methodically extend it to a Rydberg atomic MIMO system.

\subsection{LO-Free Rydberg Atomic Receiver}

The Rabi frequency $\Omega _{\text{RF}}$ is given explicitly by~\cite{QS-16}
\begin{equation}\label{Rabi_RF_LO_free}
\Omega _{\text{RF}} = \left| {  E_{\text{RF}} } \right|\frac{{{\wp _{\text{RF}}}}}{\hbar },
\end{equation}
where $\wp_{\text{RF}}$ represents the dipole moment associated with the $\left| 3 \right\rangle  \to \left| 4 \right\rangle $ transition. Eq.~(\ref{Rabi_RF_LO_free}) delineates the core principle of the LO-free Rydberg atomic receiver, indicating that the amplitude of the RF signal can be determined by reading out the separation of the AT splitting associated with the Rabi frequency $\Omega_{\text{RF}}$.

Next, we turn our attention to the modeling of wireless systems. We consider a single Tx serving as a point source, transmitting a unit complex Gaussian signal $x$ of transmit power $P_{\text{Tx}}$, obeying $ x \sim \mathcal{CN}(0, 1) $. After propagating through a complex Gaussian channel $h \sim \mathcal{CN}(0, 1)${\footnote{Here we adopt the canonical single-tap line-of-sight (LoS) channel to isolate receiver-intrinsic effects. The proposed framework remains applicable to multipath fading channels, which can be incorporated by replacing $h$ with the appropriate stochastic or tapped-delay representation.}}, the signal arrives at the vapor cell for measurement, and this process can be modeled as
\begin{equation}\label{original_y}
y = \sqrt {{P_{{\text{Rx}}}}} hx + n_{\text{ex}},
\end{equation}
where $ {P_{{\text{Rx}}}} = \frac{{{{\left| {{E_{{\text{RF}}}}} \right|}^2}}}{{2{Z_0}}} $ represents the received power, while $n_{\text{ex}}$ is the extrinsic noise. 
According to the Friis transmission equation~\cite{Friis}, the received power $P_{\text{Rx}} $ is modeled as
\begin{equation}
	{P_{{\text{Rx}}}} = {P_{{\text{Tx}}}}{G_{{\text{Tx}}}}{G_{{\text{Rx}}}}{\left( {\frac{\lambda }{{4\pi {d_{{\text{Tx-Rx}}}}}}} \right)^2},
\end{equation}
where $P_{\text{Tx}}$ denotes the transmit power, 
$G_{\text{Tx}}$ and $G_{\text{Rx}}$ are the gains of the Tx and Rx antennas, respectively, and $d_\text{Tx-Rx}$ represents the Tx-Rx link distance.
Nevertheless, for a Rydberg atomic receiver, there is no metallic antenna and hence no meaningful classical gain $G_\text{Rx}$. Instead, highly excited Rydberg atoms ``sense” the incident RF field via dipole coupling, behaving like a weakly-coupled, isotropic aperture whose effective area is determined by the quantum energy-exchange efficiency~\cite{book_antenna_theory}. Accordingly, we define the received power flux density at the Rydberg atomic receiver as
\begin{equation}
	{S_{\text{Rx}}} = \frac{{{P_{{\text{Tx}}}}{G_{{\text{Tx}}}}}}{{4\pi d_{{\text{Tx-Rx}}}^2}},
\end{equation}
which has units of ${\text{W/}}{{\text{m}}^2}$. The total received  power $P_{\text{Rx}}$ is then given by this flux multiplied by the equivalent effective aperture $A_\text{eff}$
\begin{equation}
	{P_{\text{Rx}}} = {S_{\text{Rx}}}{A_{{\text{eff}}}},
\end{equation}
where $A_{\text{eff}}$ can be formulated as
\begin{equation}\label{A_eff}
	{A_{{\text{eff}}}} = \frac{{2{Z_0}{N_{{\text{atoms}}}}{ \wp_{\text{RF}}^2 }{\omega _{{\text{RF}}}}}}{{\hbar \Gamma_{\text{FWHM}}}},
\end{equation}
where $N_\text{atoms}$ represents the number of atoms, $\omega_{{\text{RF}}} = 2 \pi f_{\text{RF}}$ is the angular frequency, and $\Gamma_{\text{FWHM}}$ denotes the full-width at half-maximum (FWHM) of the EIT spectrum. See Appendix~\ref{appendix_A_eff} for its detailed derivation.

The RF signal incident upon the vapor cell energizes the Rydberg atoms, facilitating the state transition  $\left| 3 \right\rangle  \to \left| 4 \right\rangle $. As a result, the AT splitting corresponding to the probe beam can be observed in the PD, determining the splitting interval and the Rabi frequency. 
Therefore, if we let $z$ denote our observations (physically representing the Rabi frequency) from the PD, the signaling process in (\ref{original_y}) can be recast as
\begin{equation}\label{z_LO_free}
z = \frac{{{\wp _{{\text{RF}}}}}}{\hbar }\left| {{E_{{\text{RF}}}}hx + {n_{{\text{Ry}}}}} \right|,
\end{equation}
where $n_{\text{Ry}}$ represents the noise encompassing all relevant noise sources, to be discussed in Sec.~\ref{sec_noise_sources}.

In contrast to conventional RF receivers, the LO-free architecture does not directly produce a time-domain baseband waveform; instead, it is formed over a measurement window $T_\text{m}$. This window is not universal but implementation-dependent, as it is jointly determined by the scan range, the required spectral resolution, and the averaging/SNR demands of the AT-splitting estimator. Typical Rydberg-EIT/AT measurements already span sub-millisecond to millisecond measurement windows in typical implementations~\cite{QS-34,QS-45}. Therefore, the LO-free receiver should be interpreted as a scan-and-estimate or snapshot-based architecture. In particular, since a discrete observation is produced for each scan window, its effective symbol rate, denoted by $R_s$, is fundamentally constrained by $R_s \le 1/T_\text{m}$.

During the $k$-th measurement interval $\left[ {k{T_\text{m}},\left( {k + 1} \right){T_\text{m}}} \right]$, the coupling laser is scanned across a prescribed detuning range, and the PD records the corresponding probe-transmission trace. For the discrete-time abstraction considered here, we associate an observation with a scan interval and assume that the incident RF amplitude remains approximately constant within that interval, which is given by
\begin{equation}
x\left( t \right) \approx x\left[ k \right],t \in \left[ {k{T_\text{m}},\left( {k + 1} \right){T_\text{m}}} \right].
\end{equation}
Let the coupling-laser detuning trajectory be formulated as
\begin{align}
	{\Delta _c}\left( t \right) = {\Delta _{c,\min }} + \frac{{{\Delta _{c,\max }} - {\Delta _{c,\min }}}}{{{T_m}}}\left( {t - k{T_\text{m}}} \right), \nonumber\\
	\left[ {k{T_\text{m}},\left( {k + 1} \right){T_\text{m}}} \right].
\end{align}
Then, the continuous-time PD output in the $k$-th scan window, denoted by $v_k\left( t \right) $, can be formulated as
\begin{equation}
v_k\left( t \right) = {R_{{\text{PD}}}}{P_{{\text{out}}}}\left( {{\Delta _c}\left( t \right)} \right) + n_{\text{Ry},k} \left( t \right) , t \in \left[ {k{T_\text{m}},\left( {k + 1} \right){T_\text{m}}} \right],
\end{equation}
where ${R_{{\text{PD}}}}$ denotes the PD responsivity, while $n_{\text{Ry},k} \left( t \right) $ collects the relevant noise contributions. By sampling the scan trace at $M$ detuning points, we obtain
\begin{equation}
{v_k}\left[ m \right] = {R_{{\text{PD}}}}{P_{{\text{out}}}}\left( {{\Delta _c}\left[ m \right]} \right) + {n_{{\text{Ry}},k}}\left[ m \right].
\end{equation}
The LO-free discrete observation is then defined as the output of a scan-based estimator{\footnote{A discrete observation $z \left[ k \right] $ is obtained from a complete scan interval of duration $T_\text{m}$, during which the incident RF amplitude is assumed to be approximately constant. 
		Consequently, the condition $R_s \le 1/T_\text{m}$ imposes a measurement latency of at least one scan window per observation. This reveals a basic trade-off associated with $T_\text{m}$: a larger $T_\text{m}$ generally improves the accuracy and reliability of AT-splitting estimation, but it also increases latency and reduces tolerance to rapid temporal variations.}}
\begin{equation}
z\left[ k \right] \triangleq {\Omega _{{\text{RF}}}}\left[ k \right] = \mathcal{E}\left( {{v_k}\left[ m \right]_{m = 0}^{M = 1}} \right),
\end{equation}
where $\mathcal{E}\left(  \cdot  \right)$ denotes the AT-splitting extraction procedure, such as peak-separation detection or model-based fitting of the AT splitting. Therefore, the scalar observation $z\left[ k \right]$ obtained in the LO-free architecture is precisely the estimated Rabi frequency, i.e., the model presented in (\ref{z_LO_free}). The corresponding RF electric-field can then be obtained from $z\left[ k \right]$ through the Rabi-frequency relation in \eqref{Rabi_RF_LO_free}.

\subsection{LO-Dressed Rydberg Atomic Receiver}

In contrast to the LO-free regime that is only capable of measuring the amplitude of the signal, the LO-dressed Rydberg atomic receiver can measure both the amplitude and the phase of an RF signal. Again, this technique relies on the concept of a Rydberg atom-based mixer~\cite{QS-16,QS-27,QS-28}. Briefly, an additional reference signal, also known as an LO being on-resonance with the Rydberg transition, is introduced into the vapor cell to control the down-conversion dynamics of the Rydberg atoms. This exposes the Rydberg atoms to the EIT/AT effect in order to demodulate the RF signal to be measured. The frequency difference and the phase difference between the RF signal and the LO can be optically traced. In what follows, we first elaborate on the underlying physical principles, and subsequently examine their compatibility with wireless communication systems.


As regards to the LO-dressed Rydberg atomic receiver, we specify the RF signal ${E_{{\text{RF}}}}$ and the LO signal ${E_{{\text{LO}}}}$ respectively as
\begin{subequations}
\begin{align}
&{E_{{\text{LO}}}} = {A_{{\text{LO}}}}\cos \left( {2\pi {f_{{\text{LO}}}}t + {\phi _{{\text{LO}}}}} \right), \\
&{E_{{\text{RF}}}} = {A_{{\text{RF}}}}\cos \left( {2\pi {f_{{\text{RF}}}}t + {\phi _{{\text{RF}}}}} \right),
\end{align}
\end{subequations}
where $A$, $f$, and $\phi$ denote their own amplitudes (where we define ${A_{{\text{RF}}\left( {{\text{LO}}} \right)}} = \left| {{E_{{\text{RF}}\left( {{\text{LO}}} \right)}}} \right|$), frequencies, and phases, with the frequency and phase differences given by $\Delta f = {f_{{\text{LO}}}} - {f_{{\text{RF}}}}$ and $\Delta \phi  = {\phi _{{\text{LO}}}} - {\phi _{{\text{RF}}}}$, respectively. The superposition field ${E_{{\text{total}}}}$ can be formulated as
\begin{align}\label{E_total}
{E_{{\text{total}}}} =& {E_{{\text{LO}}}} +  {E_{{\text{RF}}}} \nonumber\\
  =& {A_{{\text{LO}}}}\cos \left( {2\pi {f_{{\text{LO}}}}t + {\phi _{{\text{LO}}}} } \right) \nonumber\\
	&+ {A_{{\text{RF}}}}\cos \left( {2\pi \left( {{f_{{\text{LO}}}} - \Delta f} \right)t + {\phi _{{\text{LO}}}} - \Delta \phi } \right) \nonumber\\
	=& \sqrt {A_{{\text{LO}}}^2 + A_{{\text{RF}}}^2 + 2{A_{{\text{LO}}}}{A_{{\text{RF}}}}\cos \left( {2\pi \Delta ft + \Delta \phi } \right)}  \nonumber\\
	&\times \cos \left( {2\pi {f_{{\text{LO}}}}t + {\phi _{{\text{LO}}}}} \right) ,
\end{align}
which indicates that the non-linear response of the Rydberg atoms to the superposition field ${E_{{\text{total}}}}$ can be treated as an envelope-detection process. When we have $ {A_{{\text{RF}}}} / {A_{{\text{LO}}}} \ll 1$, the main frequency component of the superposition field's envelope is determined by the difference $\Delta f$, where the upper harmonics may be deemed negligible~\cite{QS-27,QS-28}. By performing the second-order Taylor expansion of~(\ref{E_total}) at point $\frac{{{A_{{\text{RF}}}}}}{{{A_{{\text{LO}}}}}} = 0$ with respect to $\frac{{{A_{{\text{RF}}}}}}{{{A_{{\text{LO}}}}}}$, we arrive at 
\begin{equation}\label{E_total_Taylor_expansion}
{E_{{\text{total}}}} \approx \left[ {{A_{{\text{LO}}}} + {A_{{\text{RF}}}}\cos \left( {2\pi \Delta ft + \Delta \phi } \right)} \right] \cos \left( {2\pi {f_{{\text{LO}}}}t + {\phi _{{\text{LO}}}}} \right).
\end{equation}
In (\ref{E_total_Taylor_expansion}), the resonant term, i.e., $\cos \left( {2\pi {f_{{\text{LO}}}}t + {\phi _{{\text{LO}}}}} \right)$, induces an AT splitting effect, hence reducing the peak of the EIT line, and the lower frequency term, i.e., $  {{A_{{\text{LO}}}} + {A_{{\text{RF}}}}\cos \left( {2\pi \Delta ft + \Delta \phi } \right)} $, modulates the amplitude of the resonance.

Furthermore, to provide a closed-form expression for $P_{\text{out}}$, we have to obtain $\rho_{21} \left(t \right) $ by solving the master equation in (\ref{master_equation}). Accordingly, $\Omega$ in (\ref{Hamiltonian}) is further refined as $\Omega_{{\text{total}}} =  \Omega_{\text{LO}} +  e^{\jmath \left( 2 \pi \Delta f t +\Delta \phi \right) } \Omega_{\text{RF}}  $, and ${\left| {{\Omega _{{\text{total}}}}} \right|^2} = {\Omega _{{\text{total}}}}\Omega _{{\text{total}}}^*$. However, the density matrix component $\rho_{21}$ is excessively complex and, as such, only lends itself to numerical simulations, for example, by using QuTip~\cite{QuTip}. To unveil its underlying physics model more intuitively, we derive an analytical approximation. In particular, the states $\left| 3 \right\rangle $ and $\left| 4 \right\rangle $ are metastable, exhibiting significantly extended lifetimes, resulting in spontaneous emission rates that are markedly lower than those of state $\left| 2 \right\rangle $ ($\gamma_3 \approx \gamma_4 \ll \gamma_2$). Therefore, within this approximation, it is reasonable to set $\gamma_3 = \gamma_4 =0$~\cite{QS-16}. Additionally, we consider only the resonant case where the probe laser, coupling laser, and LO are in resonance with the corresponding atomic energy levels. This implies that their detunings are $\Delta_p= \Delta_c= \Delta_{\text{LO}} = 0$. Under these circumstances, the  density matrix component $\rho_{21}$ simplifies to (the time index ``$t$" is dropped for notational conciseness)
\begin{align}\label{rho_21_LO_dress}
&{\rho _{21}}{|_{\left( {{\Delta _p},{\Delta _c},{\Delta _{{\text{LO}}}},{\gamma _3},{\gamma _4}} \right) = 0}} 
 \nonumber\\
 & = \frac{{\jmath {\gamma _2}{\Omega _p}{{\left| {{\Omega _{{\text{total}}}}} \right|}^2}}}{{\gamma _2^2{{\left| {{\Omega _{{\text{total}}}}} \right|}^2} + 2\Omega _p^2\left( {\Omega _c^2 + \Omega _p^2 + {{\left| {{\Omega _{{\text{total}}}}} \right|}^2}} \right)}}.
\end{align}
Upon substituting (\ref{rho_21_LO_dress}) into (\ref{P_out}), we obtain 
\begin{align}\label{P_out_kappa}
{P_{{\text{out}}}}\left( t \right) = {{\bar P}_0} + \kappa {\Omega _{{\text{RF}}}}\cos \left( {2\pi \Delta ft + \Delta \phi } \right),
\end{align}
where ${\bar P_0} = {P_{{\text{in}}}}{e^{ - \alpha }}{e^{\alpha \Lambda \left( {{\Omega _{{\text{LO}}}},\Gamma } \right)}}$ represents the average intensity of the probe laser. Furthermore, $\alpha  = k_pL{C_0}\bar A$ models the absorption coefficient of the probe laser, $\bar A = {{{\gamma _2}{\Omega _p}}}/\left( {{\gamma _2^2 + 2\Omega _p^2}}\right) $ can be understood as the amplitude of the three-level EIT spectrum, $\kappa  = \alpha {{\bar P}_0}{\kappa _\rho }$ is a key intrinsic coefficient characterizing the slope induced when the AT splitting of the EIT peak shifts the on-resonance point to the slope of each EIT line, ${\kappa _\rho } = {{\partial \Lambda \left( {{\Omega _{{\text{LO}}}},\Gamma } \right)}}/{{\partial {\Omega _{{\text{LO}}}}}}$ represents the intrinsic gain coefficient associated with the density matrix component $\rho_{21}$, $\Gamma  = {\Omega _p}\sqrt {2\left( {\Omega _c^2 + \Omega _p^2} \right)/\left( {2 \Omega _p^2 + \gamma _2^2} \right)} $ represents the half-width at half-maxima (HWHM) of a four-level system, as well as the function $\Lambda \left( {a,b} \right) = {{{b^2}}}/\left( {{{b^2} + {a^2}}}\right) $  is a normalized Lorentzian function with $a$ as the variable and $b$ representing the HWHM, where $2b$ corresponds to the full-width at half-maxima (FWHM). See Appendix~\ref{derivation_P_out}  for detailed derivations.

Then, the LO-dressed Rydberg atomic receiver measures an RF signal in form of an optical readout, represented by
\begin{align}\label{P_readout_LO}
{P_{{\text{readout}}}}\left( t \right) &= {P_{{\text{out}}}}\left( t \right) - {{\bar P}_0} \nonumber\\
&= \left| {P\left( {\Delta f} \right)} \right|\cos \left( {2\pi \Delta ft + \Delta \phi } \right),
\end{align}
where $\left| {P\left( {\Delta f} \right)} \right| $ is the amplitude of the single-sided Fourier spectrum at the frequency $\Delta f$ when performing the Fourier transform over $P_\text{out} \left( t\right) $, whose physical meaning corresponds to the power intensity of the output probe laser observed. The Rabi frequency of the RF signal is then given by
\begin{equation}\label{Rabi_freq_LO_readout}
{\Omega _{{\text{RF}}}} = \frac{{\left| {P\left( {\Delta f} \right)} \right|}}{{\left| \kappa  \right|}},
\end{equation}
which thus provides us with an RF signal amplitude of ${A_{{\text{RF}}}} = {\hbar {\Omega _{{\text{RF}}}}} / { {\wp _{{\text{RF}}}}}$.

Eqs.~(\ref{P_readout_LO}) and (\ref{Rabi_freq_LO_readout}) are at the heart of the LO-dressed Rydberg atomic receiver. In sharp contrast to the LO-free context, the LO-dressed one concentrates on measuring the amplitude of the single-side Fourier spectrum of $P_{\text{out}}$, i.e., $\left| {P\left( {\Delta f} \right)} \right| $, at the frequency $\Delta f$. Then, the phase difference $\Delta \phi$ can be extracted from $P_{\text{out}}$ by employing a homodyne detector~\cite{LIA,LIA-2,QS-16}, retrieving the RF signal phase according to $\phi_{{\text{RF}}} = \Delta \phi - \phi_{{\text{LO}}}$. Thus, the RF signal can be accurately recovered. Additionally, both $\left| {P\left( {\Delta f} \right)} \right| $ and ${\left| \kappa  \right|}$ can be directly attained from the optical spectrum. This technique efficiently reduces the electric-field measurement to an optical-frequency readout, while ensuring its International System of Units (SI) traceability~\cite{QS-22}.

Then, we turn our attention to the conversion process of the RF signal within this atomic system. Before proceeding, we define the key noise terms that arise throughout the context. Specifically, let ${n_{{\text{ex}}}}$ denote the extrinsic noise; ${n_{{\text{PSN}}}}$ the photon shot noise (PSN); ${n_{{\text{QPN}}}}$ the quantum projection noise (QPN); and ${n_{{\text{TN}}}}$ the thermal noise. Their respective variances are denoted by $\sigma _{{\text{ex}}}^2$, $\sigma _{{\text{QPN}}}^2$, ${\sigma _{{\text{PSN}}}^2}$, and ${\sigma _{{\text{TN}}}^2}$, which will be used throughout our discussion, and a detailed quantitative treatment will be provided later in Sec.~\ref{sec_noise_sources}. 

The probe laser beam is received by a PD and converted into an output photocurrent signal. The relationship between the alternating current (AC) component of the probe laser power, i.e., $\left| \kappa  \right|{\Omega _{{\text{RF}}}}$, and the photocurrent $I_{\text{AC}}$ is given by 
\begin{align}
{I_{{\text{AC}}}} &= D\left( {\left| {P\left( {\Delta f} \right)} \right| + {n_{{\text{PSN}}}}} \right) \nonumber\\
 &= D\left[ {\left| \kappa  \right|\frac{{{\wp _{{\text{RF}}}}}}{\hbar }\left( {\sqrt {{P_{{\text{Rx}}}}}  + {n_{{\text{ex}}}}} \right) + {n_{{\text{PSN}}}}} \right] \nonumber\\
 &= D\left| \kappa  \right|\frac{{{\wp _{{\text{RF}}}}}}{\hbar }\sqrt {{P_{{\text{Rx}}}}}  + D\left| \kappa  \right|\frac{{{\wp _{{\text{RF}}}}}}{\hbar }{n_{{\text{ex}}}} + D{n_{{\text{PSN}}}},
\end{align}
where $D$ is the photodiode's responsivity with units of A/W. Consequently, the instantaneous power associated with the AC component in the photocurrent output from the PD is given by  
\begin{align}
{P_{{\text{eAC}}}} &= I_{{\text{AC}}}^2{R_{\text{L}}} \nonumber\\
 &= {R_{\text{L}}}{D^2}{\kappa ^2}\frac{{\wp _{{\text{RF}}}^2}}{{{\hbar ^2}}}{P_{{\text{Rx}}}} + {R_{\text{L}}}{D^2}{\kappa ^2}\frac{{\wp _{{\text{RF}}}^2}}{{{\hbar ^2}}}\sigma _{{\text{ex}}}^2 + {R_{\text{L}}}{D^2}\sigma _{{\text{PSN}}}^2,
\end{align}
where $R_{\text{L}}$ is the output impedance of the PD. Then, the output power of the low-noise amplifier (LNA) is given by
\begin{align}\label{P_LNA}
	{P_{{\text{LNA}}}} =& {G_{{\text{LNA}}}}{P_{{\text{eAC}}}} + \sigma _{{\text{TN}}}^2 \nonumber\\
	=& {G_{{\text{LNA}}}}{R_{\text{L}}}{D^2}{\kappa ^2}{ \frac{{\wp _{{\text{RF}}}^2}}{{{\hbar ^2}}}}{P_{{\text{Rx}}}} \nonumber\\
	&+ {G_{{\text{LNA}}}}{R_{\text{L}}}{D^2}{\kappa ^2}\frac{{\wp _{{\text{RF}}}^2}}{{{\hbar ^2}}}\sigma _{{\text{ex}}}^2 + {G_{{\text{LNA}}}}{R_{\text{L}}}{D^2}\sigma _{{\text{PSN}}}^2 + \sigma _{{\text{TN}}}^2.
\end{align}
where ${{G_{{\text{LNA}}}}}$ denotes the gain of the photocurrent after processing by the subsequent circuits (such as a low-noise amplifier, LNA).
Given a complex Gaussian  signal transmitted from the Tx, the observation ${z^{{\text{LO}}}}$ at the PD after encountering the channel $h$ is given by{\footnote{The complex RF envelop $x $ at carrier $f_\text{RF}$ is shifted to a low intermediate frequency at $\Delta f = f_\text{LO} - f_\text{RF}$. After low-pass filtering and demodulation at $\Delta f$, this yields a complex baseband signal proportional to $x$, which leads directly to the model given in~(\ref{z_LO_dress}).}}
\begin{equation}\label{z_LO_dress}
{z^{{\text{LO}}}} = \sqrt {{P_{{\text{Rx}}}}{G_{{\text{LNA}}}}{R_{\text{L}}}} D\left| \kappa  \right|\frac{{{\wp _{{\text{RF}}}}}}{\hbar }hx + {n_{{\text{Ry,LO}}}},
\end{equation}
where $n_{{\text{Ry,LO}}}$ models the noise encompassing all relevant noise sources, as will be discussed in Sec.~\ref{sec_noise_sources}.

\begin{figure*}[t]
	\centering
	\includegraphics[width=0.92\textwidth]{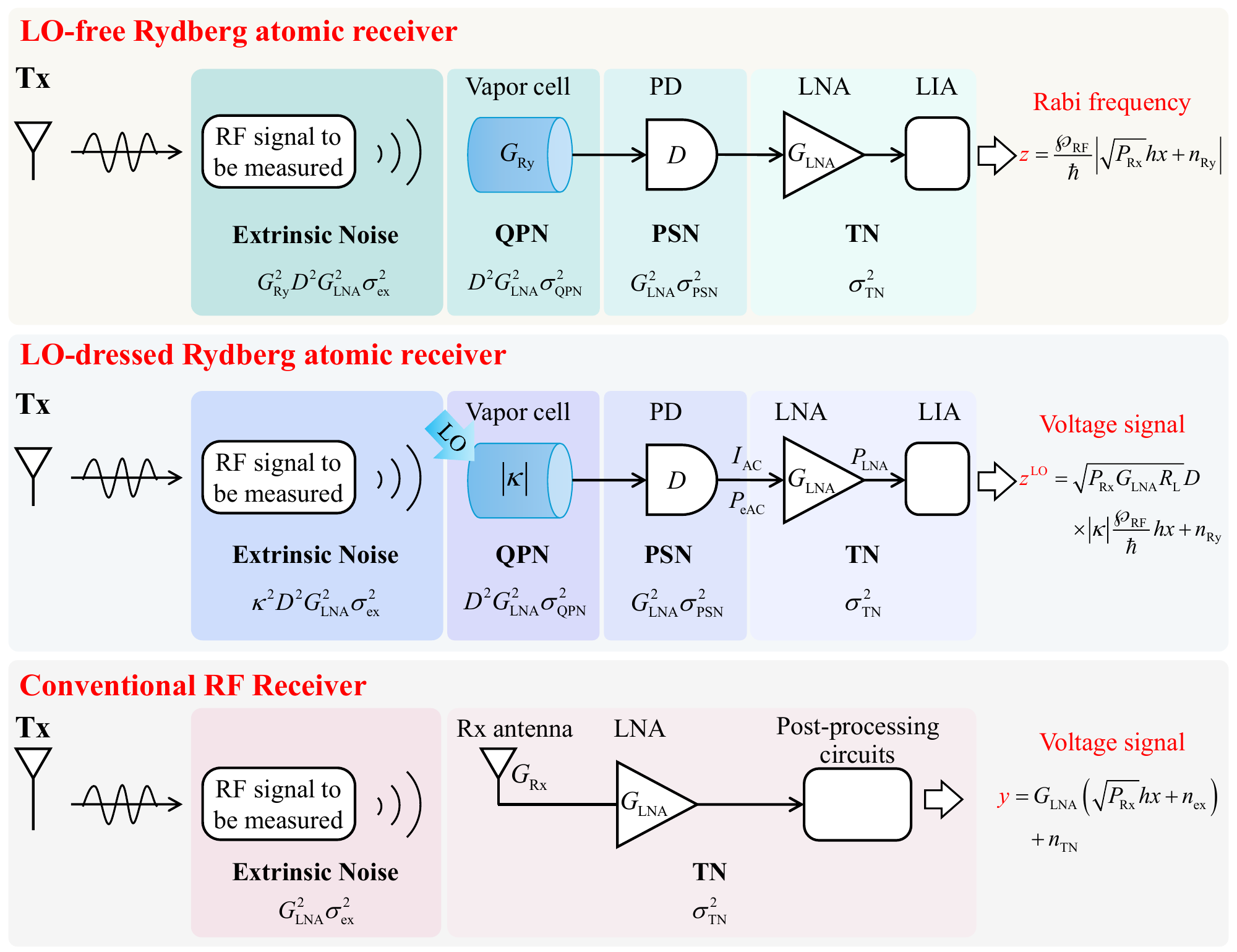}
	\caption{Noise modeling for the LO-free and the LO-dressed Rydberg atomic receivers as well as the conventional RF receiver.} \label{noise_model}
		\vspace{-1em}
\end{figure*}

\subsection{Noise Modeling}\label{sec_noise_sources}
Regardless, whether the Rydberg atomic receiver operates in an LO-free (direct AT-splitting readout) or an LO-dressed (mixer/heterodyne) configuration, the noise sources of the overall chain remain essentially the same. In both architectures, as illustrated in Fig.~\ref{noise_model}, the detected RF signal first suffers from an extrinsic, black-body radiation (BBR)-induced noise floor that couples into the Rydberg manifold as an electric-field fluctuation. Then, super-imposed on this external term are three intrinsic noise sources: QPN, PSN, and TN. While many of these noise sources appear in both configurations, their underlying physical mechanisms differ subtly. In what follows, we first clarify these key noise sources, then we elaborate on their relative impact in each configuration.

\subsubsection{Noise Sources}
To put the discussion of noise sources on a solid foundation, we introduce the noise-equivalent field (NEF)~\cite{QS-32-A,QS-34}
\begin{equation}
	{\text{NEF}} = \frac{{{{\left| {{E_{{\text{RF}}}}} \right|}_{\min }}}}{{\sqrt B }} ,
\end{equation}
which represents a bandwidth-normalized noise floor expressed in the receiver's natural ``field" units. A smaller NEF corresponds to a lower minimum detectable field, indicating higher receiver sensitivity. 
\begin{itemize}
	\item \textbf{Extrinsic Noise:} 
	In free space, the external noise originates from the thermal BBR{\footnote{Environmental BBR is treated as a free-space input characterized by the Planck/Callen-Welton spectral factor~\cite{QS-32-A,2010_Microwave_Mag,website-radiometers}. As such, it is common to classical and quantum receivers and enters at the RF input before device-specific transduction. We therefore reference BBR to the input field and propagate it through each readout chain (LO-free, LO-dressed, and conventional).}} at the ambient physical temperature and vacuum fluctuations of the RF field.
Taking both effects into account, the external noise equivalent field ${\text{NE}}{{\text{F}}_{{\text{ex}}}}$ is explicitly modeled as~\cite{QS-32-A}
	\begin{equation}
		{\text{NE}}{{\text{F}}_{{\text{ex}}}} = \sqrt {\frac{{16\pi f_{{\text{RF}}}^2}}{{3{\epsilon _0}{c^3}}}\Theta \left( {{f_{{\text{RF}}}},T} \right)} ,
	\end{equation}
	where ${\Theta \left( {{f_{{\text{RF}}}},T} \right)}$ is a modified version of the Callen-Welton law~\cite{QS-32-26}
	\begin{align}
		\Theta \left( {{f_{{\text{RF}}}},T} \right) = \hbar {f_{{\text{RF}}}}\left\{ \begin{gathered}
			\frac{1}{2}{n_{{\text{th}}}}\left( {{f_{{\text{RF}}}},T} \right) + \frac{1}{2}, \ {\text{homodyne}}, \hfill \\
			2{n_{{\text{th}}}}\left( {{f_{{\text{RF}}}},T} \right) + 1, \  \ {\text{heterodyne}}, \hfill \\ 
		\end{gathered}  \right.
	\end{align}
	with
	\begin{equation}
		{n_{{\text{th}}}}\left( {{f_{{\text{RF}}}},T} \right) = \frac{1}{{\exp \left( {\frac{{\hbar {f_{{\text{RF}}}}}}{{{k_\text{B}}T}}} \right) - 1}},
	\end{equation}
	where $k_\text{B}$ denotes the Boltzmann constant.
	The $\text{NEF}_{\text{ex}}$ sets a fundamental noise floor that no free-space receiver can overcome.
	Note that the $\text{NEF}_{\text{ex}}$ is quoted in ${{\text{V}}/{\text{m}}/\sqrt {{\text{Hz}}} }$, while system-level calculations always require noise power spectral density in $ {\text{W}}/\sqrt {{\text{Hz}}}  $. Given an effective aperture $A_\text{eff}$ and free-space impedance $Z_0$, the extrinsic noise power $\sigma _{{\text{ex}}}^2$  over a bandwidth $B$ is given by
	\begin{equation}
		\sigma _{{\text{ex}}}^2 = \frac{{{\text{NEF}}_{{\text{ex}}}^2}}{{2{Z_0}}}{A_{{\text{eff}}}}B.
	\end{equation}
	\begin{remark}
		\color{black}
In this work, $\mathrm{NEF}_{\mathrm{ex}}$ is adopted as a system-level input-referred metric that captures the impact of external electromagnetic fluctuations, including ambient BBR and microwave vacuum fluctuations~\cite{chen-Rydberg2}. This abstraction provides a tractable basis for evaluating receiver-level performance metrics such as SNR, mutual information, and SER.
At the microscopic atomic-physics level, BBR may also affect the EIT response through temperature-dependent decay and dephasing rates, resulting in coherence damping and linewidth broadening~\cite{PRL_2026}. Extending the master-equation model to include microscopic BBR-induced decoherence effects, together with a corresponding re-optimization of the receiver parameters, represents a meaningful future research direction beyond the scope of the present framework.
	\end{remark}
	
	\item (Intrinsic Noise) \textbf{QPN:} QPN, often referred to as \textit{atomic shot noise}, originates from the fundamentally probabilistic collapse of each atom's wavefunction during measurement. When $N_\text{atoms}$ mutually uncorrelated atoms are employed, the root-mean-square phase uncertainty is characterized by $1/\sqrt{N_\text{atoms}}$, establishing the SQL on the field detection. For a Rydberg atomic receiver, this translates into an SQL electric-field amplitude $E_\text{SQL}$, which is given by~\cite{QS-34}
	\begin{equation}\label{SQL}
		\frac{{{E_{{\text{SQL}}}}}}{{\sqrt {{\text{Hz}}} }} = \frac{\hbar }{{{\wp _{{\text{RF}}}}\sqrt {{N_{{\text{atoms}}}} T_2 } }},
	\end{equation}
	where $T_2$ denotes the coherence time of the EIT process.	The power of QPN, denoted by $\sigma_\text{QPN}$, is formulated as
	\begin{equation}
		\sigma _{{\text{QPN}}}^2 = {\left( {\frac{{{E_{{\text{SQL}}}}}}{{\sqrt {{\text{Hz}}} }}} \right)^2}\frac{{{A_{{\text{eff}}}}B}}{{2{Z_0}}}.
	\end{equation}
	
	\item (Intrinsic Noise) \textbf{PSN:} The probe laser having an optical power of $\left|  P_\text{out} \right|$ impinges on a PD with quantum efficiency~${\eta _{{\text{eff}}}} \in \left[ 0,1 \right] $. The average photo-electron rate is $\frac{{{ \left|  P_\text{out} \right| }}}{{2\pi \hbar {f_p}}}  $, where  $f_p$ denotes the probe laser frequency, yielding the average photocurrent of
	\begin{equation}\label{average_photocurrent}
		\bar I =  \frac{  { \eta _{\text{eff}} e } }{{2\pi \hbar {f_{{\text{p}}}}}} \left|  P_\text{out} \right| .
	\end{equation}
	The power of PSN, denoted by $\sigma_{\text{PSN}}^2$, is formulated as
	\begin{equation}\label{sigma2_PSN}
		\sigma _{{\text{PSN}}}^2 = 2eB\bar I.
	\end{equation}
	
	\item (Intrinsic Noise) \textbf{TN:} TN, also referred to as the Johnson-Nyquist noise, is typically induced by the random motions of charge carriers in resistive elements and enters the receiver through different components depending on the front-end topology. In the case of a trans-impedance amplifier (TIA) front-end, the dominant source is the feedback resistor $R_\text{L}$ of the TIA. The Johnson noise from this resistor manifests itself at the TIA output as a mean-square voltage of $4k_\text{B}TR_\text{L}B$, corresponding to an output-referred power of $4k_\text{B}TB$ when $R_\text{L} = 1~\text{Ohm}$~\cite{TIA-handbook}. For the LNA front-end, the TN from a matched source is $k_\text{B}TB$. After amplification by the first-stage LNA, this noise is scaled by the LNA noise factor $F$, resulting in a total output noise power of $Fk_\text{B}TB$. Accordingly, we model the TN power as
	\begin{equation}
\sigma _{{\text{TN}}}^2 = \left\{ \begin{gathered}
	4{k_\text{B}}TB,~\left( {{\text{TIA front-end}}} \right), \hfill \\
	F{k_\text{B}}TB,~\left( {{\text{LNA front-end}}} \right). \hfill \\ 
\end{gathered}  \right.
	\end{equation}
	
\end{itemize}

\subsubsection{LO-Free Rydberg Atomic Receiver}
For the LO-free Rydberg atomic receiver that relies on the direct readout of AT splitting, we introduce the small-signal transduction gain of
\begin{equation}
{G_{{\text{Ry}}}} = \frac{{\partial {P_{{\text{out}}}}}}{{\partial {E_{{\text{RF}}}}}},
\end{equation}
which characterizes the differential change in probe laser transmission $P_{{\text{out}}}$ per unit incident RF electric field $E_{{\text{RF}}} $. Although the LO-free architecture does not employ a mixing process, $G_{{\text{Ry}}}$ is indispensable for a complete noise analysis: it allows all downstream noise sources (photon-shot, BBR, electronic, etc.) to be pursued back to the RF input using the standard Friis cascade formalism~\cite{book_Friis_noise}. Therefore, the noise power of $n_\text{Ry}$ in ~(\ref{z_LO_free}), denoted by $\sigma _{{\text{Ry}}}^2$, is expressed as
\begin{align}\label{noise_power_LO_free}
\sigma _{{\text{Ry}}}^2 = G_{{\text{Ry}}}^2{D^2}G_{{\text{LNA}}}^2\sigma _{{\text{ex}}}^2 + {D^2}G_{{\text{LNA}}}^2\sigma _{{\text{QPN}}}^2 + G_{{\text{LNA}}}^2\sigma _{{\text{PSN}}}^2 + \sigma _{{\text{TN}}}^2.
\end{align}

\subsubsection{LO-Dressed Rydberg Atomic Receiver}
Similarly, the noise power in~(\ref{z_LO_dress}), denoted by $\sigma _{{\text{Ry,LO}}}^2$, is expressed as
\begin{equation}\label{noise_model_LO_dressed}
\sigma _{{\text{Ry,LO}}}^2 = {\kappa ^2}{D^2}G_{{\text{LNA}}}^2\sigma _{{\text{ex}}}^2 + {D^2}G_{{\text{LNA}}}^2\sigma _{{\text{QPN}}}^2 + G_{{\text{LNA}}}^2\sigma _{{\text{PSN}}}^2 + \sigma _{{\text{TN}}}^2.
\end{equation}
We remark here that $G_{\text{Ry}}$  is indispensable whenever we push optical-domain noise back to the RF input, but its absolute value is modest because the LO-free architecture operates on the natural slope of an EIT doublet. Regarding an LO-dressed Rydberg atomic receiver, the same microscopic derivative is evaluated at a bias set by a strong LO, yielding a coefficient $\left| \kappa\right| $ that can be one to two orders of magnitude higher and that therefore becomes the primary figure of merit for sensitivity engineering~\cite{QS-28}.

\section{SNR Performance and Distortion Effect}\label{section_distortion_effect}

Based on the universal models for LO-free and LO-dressed Rydberg atomic receivers presented in~(\ref{z_LO_free}) and (\ref{z_LO_dress}), this section derives the closed-form expressions for their SNR. Next, we examine the distortion effects intrinsic to each receiver, including distortion effect, in order to delineate their practical operating ranges.
\subsection{SNR Performance}
\subsubsection{LO-Free Rydberg Atomic Receiver}

In contrast to the conventional RF receiver that evaluates the link quality solely through the ratio between received power and front-end noise power, the LO-free Rydberg atomic receiver does not reconstruct the incident field as a linear voltage. Instead, the RF field drives a Rydberg-Rydberg transition with Rabi frequency $\Omega_{{\text{RF}}}$ and the information we finally read out is the AT splitting interval as in (\ref{AT-interval}). The sharpness with which this interval can be resolved is determined by the EIT linewidth $\Gamma_{\text{FWHM}}$. Therefore, the SNR must account, besides the link budget, for how clearly the two AT peaks separate. To capture this effect, we introduce a dimensionless $\mathcal{R}$ for measuring the RF coupling strength in units of the intrinsic EIT linewidth
\begin{equation}\label{ratio_R}
\mathcal{R} \equiv \frac{{{\Omega _{{\text{RF}}}}}}{{{\Gamma _{{\text{FWHM}}}}}}.
\end{equation}
The Fisher-information analysis in Appendix~\ref{appendix_SNR_LO_free} shows that the SNR is required to be multiplied by
\begin{equation}
G\left( \mathcal{R} \right) = \frac{{{\mathcal{R}^2}}}{{1 + \frac{1}{{2{\mathcal{R}^2}}}}},
\end{equation}
which is an information penalty that quantifies the resolvability loss of the AT splitting.

Then, the SNR expression of an LO-free Rydberg receiver reads
\begin{equation}\label{SNR_LO_free}
{\text{SN}}{{\text{R}}_{{\text{Ry}}}} = \frac{{{P_{{\text{Rx}}}}{{\left| h \right|}^2}}}{{\sigma _{{\text{Ry}}}^2}}{\left( {\frac{{{\wp _{{\text{RF}}}}}}{{\hbar {\Gamma _{{\text{FWHM}}}}}}} \right)^2}G\left( \mathcal{R} \right).
\end{equation}
One may refer to Appendix~\ref{appendix_SNR_LO_free} for its derivation. The SNR expression in (\ref{SNR_LO_free}) integrates the wireless link budget with the quantum-optical response of the LO-free Rydberg atomic receiver. More specifically, when $ \mathcal{R} \gg 1$, the SNR reduces to the Friis-type link SNR scaled by a constant conversion gain. By letting $\text{SNR}_\text{Ry} = 1$, we can obtain the sensitivity of the LO-free Rydberg atomic receiver, which is given by 
\begin{equation}\label{LO_free_sensitivity}
E_{\min }^{{\text{strong}}} = \frac{{\hbar {\Gamma _{{\text{EIT}}}}}}{{{\wp _{{\text{RF}}}}}}{\left( {\frac{{2{Z_0}\sigma _{{\text{Ry}}}^2}}{{{A_{{\text{eff}}}}{{\left| h \right|}^2}}}} \right)^{\frac{1}{4}}}.
\end{equation}
By contrast, when $ \mathcal{R} \ll 1$, the factor $G\left( \mathcal{R} \right)$ forecasts the precipitous SNR drop that accompanies an ambiguous AT splitting interval.

\subsubsection{LO-Dressed Rydberg Atomic Receiver}
According to the analysis in (\ref{P_LNA}), we can extract both the received signal power and the noise power, hence yield the SNR expression for the LO-dressed Rydberg atomic receiver
\begin{equation}\label{SNR_LO_dressed}
{\text{SN}}{{\text{R}}_{{\text{Ry,LO}}}} = \frac{{{G_{{\text{LNA}}}}{R_{\text{L}}}{D^2}{\kappa ^2}\frac{{{\wp_{\text{RF}} ^2}}}{{{\hbar ^2}}}{P_{{\text{Rx}}}}{{\left| h \right|}^2}}}{{\sigma _{{\text{Ry,LO}}}^2}}.
\end{equation}

\begin{figure}[t]
	\centering
	\includegraphics[width=0.5\textwidth]{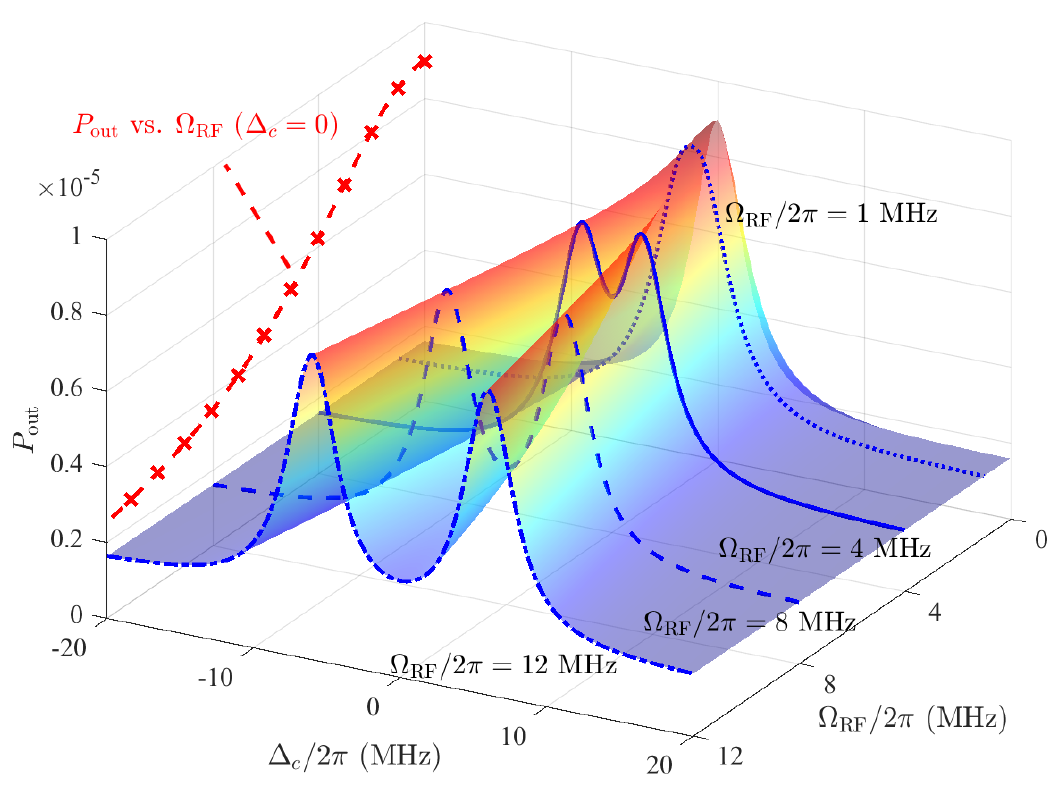}
	\caption{The probe laser transmission $P_{\text{out}}$ versus the coupling detuning $\Delta_c$ and RF Rabi frequency $\Omega_{\text{RF}}$.} \label{LO_free_contour_3d}
	\vspace{-1em}
\end{figure}

\subsection{Distortion Effect}
This subsection characterizes the distortion effects associated with these two types of Rydberg atomic receivers by simulation results, thereby highlighting their respective operating conditions, when used for wireless systems. Before we proceed, it is essential to clarify some critical parameters related to our simulation results. Unless otherwise specified for a particular type of Rydberg atomic receiver, these parameters apply to both types of receivers.

We employ the QuTiP toolkit~\cite{QuTip} for simulating the quantum system considered, where the parameters are taken from~\cite{QS-12,QS-16,QS-28} to ensure their rationale. To be specific, the vapor cell is $L = 1~\text{cm}$ long and comprises ground state atoms at a total density of $N_0 = 4.89 \times 10^{10}~{\text{cm}}^{-3}$.
The four-level energy system illustrated in Fig.~\ref{system_model}(a) is realized by taking into  account the following four states $\left| 1 \right\rangle  \to \left| 2 \right\rangle  \to \left| 3 \right\rangle  \to \left| 4 \right\rangle $ in a Cs atom: $6{S_{1/2}} \to 6{P_{3/2}} \to 47{D_{5/2}} \to 48{P_{3/2}}$, in which the Rydberg state $47{D_{5/2}}$ has an inverse lifetime of $\gamma_3 /  2\pi  = 3.9~\text{kHz}$, and the Rydberg state $48{P_{3/2}}$ has an inverse lifetime of $\gamma_4 /  2\pi  = 1.7~\text{kHz}$. For the two lowest states of $\left| 1 \right\rangle  $ and $ \left| 2 \right\rangle$, the lifetimes are $\gamma_1 = 0$ and $\gamma_2 /  2\pi  = 5.2~\text{MHz}$, respectively. {The dephasing rate $\gamma_{ij}$ can be calculated by ${\gamma _{ij}} = \left( { {\gamma _i} + {\gamma _j} } \right)/2$.}
The dipole moment associated with a $\left| 3 \right\rangle  \to \left| 4 \right\rangle $ transition is $\wp_{{\text{RF}}} = -1443.459 e a_0$, and $\wp_{12} = \left( 2.5 e a_0 \right) ^2$ for  $\left| 1 \right\rangle  \to \left| 2 \right\rangle $ transition, with $e = 1.6\times 10^{-19}~\text{C}$ representing the elementary charge and $a_0 = 5.2\times 10^{-11}~\text{m}$ for the Bohr radius. The carrier frequency of the RF signal is $f_{\text{RF}} = 6.9~\text{GHz}$. For the probe laser, 
the wavelength is $\lambda_p= 852~{\text{nm}}$, the laser intensity is $20.7~\mu \text{W}$, and the $1/{e^2}$ beam diameter is $0.76~\text{mm}$, yielding the Rabi frequency of $\Omega_p / 2\pi  = 8~\text{MHz}$. For the coupling laser, the wavelength is $\lambda_c= 510~{\text{nm}}$, the laser intensity is $17~\text{mW}$, the $1/{e^2}$ beam diameter is $1.95~\text{mm}$, and the Rabi frequency is  $\Omega_c /  2\pi  = 1~\text{MHz}$. The permittivity in vacuum is $\epsilon_0 = 8.854\times10^{-12}~\text{F/m}$, and the impedance in free space is given by $Z_0 = 377~\text{Ohm}$.
Regarding the LO-free Rydberg atomic receiver, the Rabi frequency of the RF field is $\Omega_{{\text{RF}}} /  2\pi  = 6~\text{MHz}$. For the LO-dressed system, the Rabi frequency of the RF field is $\Omega_{{\text{RF}}} /  2 \pi  = 20~\text{kHz}$ and $\Omega_{{\text{LO}}} /  2\pi  = 4.23~\text{MHz}$ for the LO field. The bandwidth of the atomic system is $B = 100~\text{kHz}$, while the frequency difference is given by $\Delta f = 15~\text{kHz}$, comfortably below the $3\text{-dB}$ bandwidth{\footnote{In the LO-dressed readout, the atomic-optical chain acts as a low-pass stage with 3-dB bandwidth $B$. The information band $\left[ {\Delta f - {B_{{\text{sig}}}},\Delta f + {B_{{\text{sig}}}}} \right]$ (with $B_\text{sig}$ denoting signal bandwidth) must lie inside the passband, i.e., ${B_{{\text{sig}}}} \le B - \Delta f$.}}. The remaining parameters are specified as follows: $G_{\text{LNA}} = 20~\text{dB}$, $R_\text{L} = 50~\text{Ohm}$, $D = 0.55~\text{A/W}$, $T= 290~\text{Kelvin}$, and $\eta_{{\text{eff}}} = 0.5$.

We then justify the rationale of the ``independent-atom" assumption. The ``independent-atom" master equation is determined not only by the atomic density $N_0$ alone, but by the instantaneous Rydberg excitation density $n_R = \eta_R N_0$ and the blockade figure of merit, denoted by ${N_b} \triangleq {n_R}\frac{{4\pi r_b^3}}{3}$. In this expression, ${r_b} = {\left( {\frac{{\left| {{C_6}} \right|}}{{\hbar {\Gamma _{{\text{EIT}}}}}}} \right)^{1/6}}$ is the van-der-Waals blockade radius, while $\eta_R \lesssim \left( \Omega_p /\Omega_c\right) ^2$ represents the Rydberg fraction, and $C_6$ is the van-der-Waals coefficient~\cite{QS-67-booklet}. In light of~\cite[Sec.~V]{QS-67-booklet}, the criterion for negligible atom interactions is $N_b \ll 1$. Using $B= 100~\text{kHz}$, the van-der-Waals blockage radius for the $47D_{5/2}$ state is approximately $13~\mu\text{m}$, giving a blockage volume of $\frac{{4\pi r_b^3}}{3}\approx 1.2\times10^{-13}~\text{m}^3$~\cite{ARC} at the atomic density of $N_0 = 4.89\times 10^{10}~\text{cm}^{-3}$ and weak-probe Rydberg fraction $\eta_R \lesssim 3\times 10^{-4}$. As a result, the average number of Rydberg excitation per blockade sphere is given by ${N_b} = \eta_R N_0 \frac{{4\pi r_b^3}}{3} \approx 0.02 \ll 1$. Hence atom-atom interactions can be safely neglected.

\subsubsection{LO-Free Rydberg Atomic Receiver}

\begin{figure}[t]
	\centering
	\includegraphics[width=0.47\textwidth]{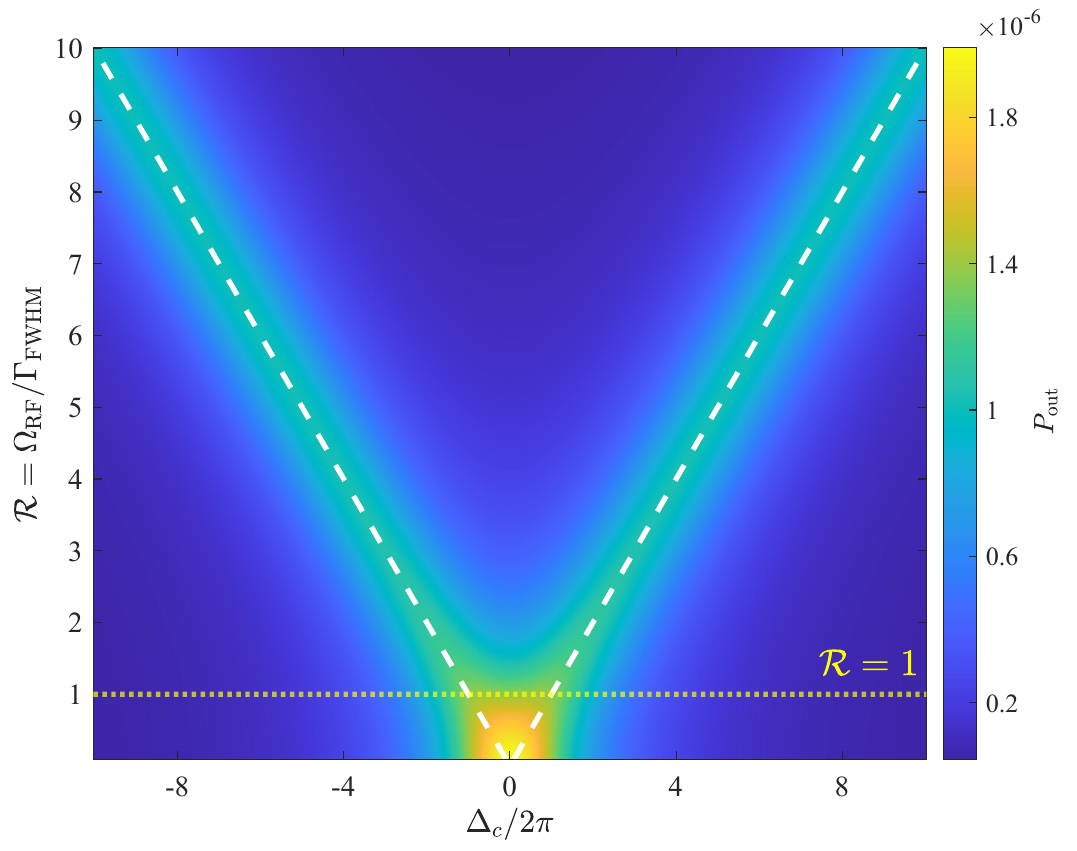}
	\caption{The probe laser transmission heatmap as a function of the coupling detuning $\Delta_c / 2\pi$ and the ratio $\mathcal{R}$.} \label{LO_free_ratio_R}
		\vspace{-1em}
\end{figure}

As regards to the LO-free Rydberg atomic receiver, the distortion may arise from ambiguous observations when the Rabi frequency $\Omega_{{\text{RF}}}$ is weak, and hence the AT splitting interval cannot be accurately identified. 
Fig.~3 illustrates the dependence of the AT splitting interval on the RF‐driving strength $\Omega_{{\text{RF}}}$. When $\Omega_{{\text{RF}}}$ is large compared to the intrinsic linewidth, two well‐resolved transmission peaks appear at $\Delta_c \approx \pm \Omega_{{\text{RF}}}/2$.  As $\Omega_{{\text{RF}}}$ decreases towards the same order as $\Gamma_\text{FWHM}$, the two peaks merge into a single broadened feature and the splitting becomes unobservable. This unresolved regime directly underlies the distortion in the LO-free context: below a critical RF field strength the AT splitting interval can no longer be clearly identified.
If we fix $P_{\text{out}}$ in Fig.~\ref{LO_free_contour_3d} and scan $\Delta_c$, we obtain the trend of the AT splitting interval upon varying the Rabi frequency $\Omega_{{\text{RF}}}$, as shown in Fig.~\ref{LO_free_ratio_R}, which, actually, recasts this distortion behavior leveraging the dimensionless ratio $\mathcal{R}$ presented in~(\ref{ratio_R}). Specifically, the white dashed lines in Fig.~\ref{LO_free_ratio_R} trace the loci $\Delta_c/ 2\pi = \pm \mathcal{R}$ of the AT peaks, and the horizontal yellow dotted line marks the threshold $\mathcal{R} = 1$. For $\mathcal{R}<1$, the transparency window at $\Delta_c=0$ remains single‐peaked, indicating complete merging of the AT splitting. Only when $\mathcal{R}>1$ do the two peaks emerge and they become separated linearly upon increasing $\mathcal{R}$. In this case, we calculate the sensitivity of the LO-free Rydberg atomic receiver to be $38.15~\mu\text{V}/\text{cm}/\sqrt{\text{Hz}}$ according to (\ref{LO_free_sensitivity}). This value is consistent with the magnitude reported in the existing literature~\cite{QS-3,QS-19}, confirming the validity of our proposed SNR expression. Hence, the dimensionless ratio $\mathcal{R}$ serves as a compact figure of merit: achieving $\mathcal{R}\gg 1$ is a necessary condition for clear AT splitting and high‐fidelity LO-free Rydberg sensing.

\begin{figure}[t]
	\centering
	\includegraphics[width=0.49\textwidth]{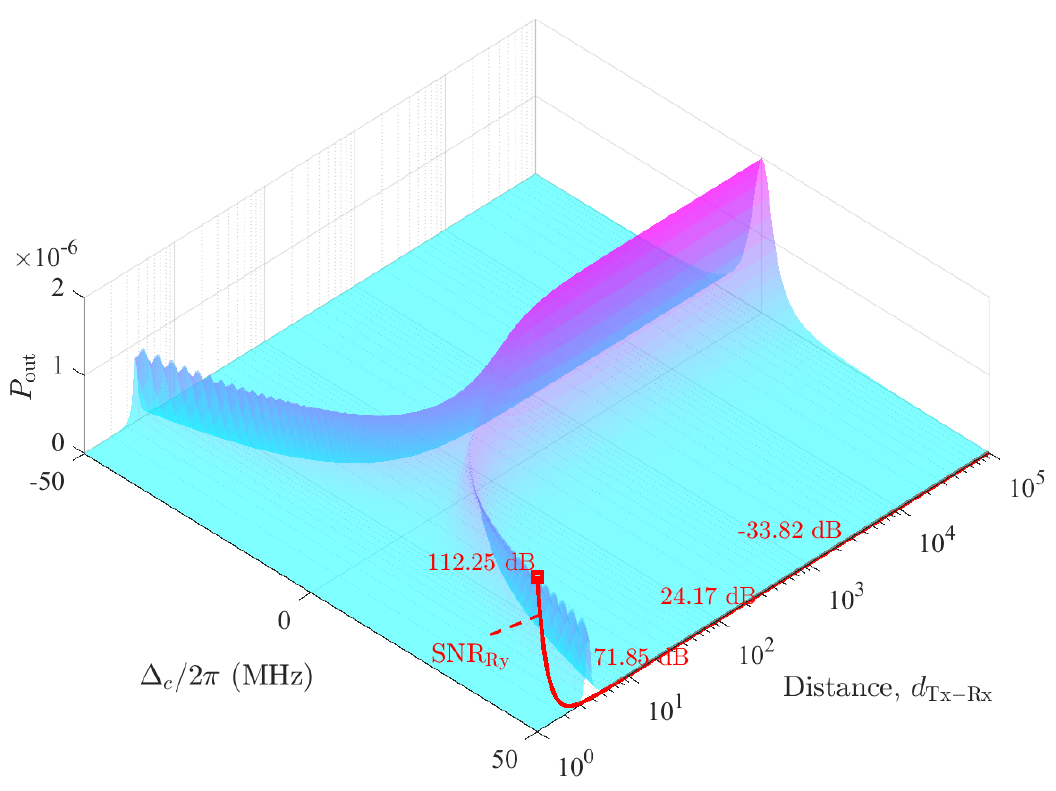}
	\caption{The probe laser transmission $P_\text{out}$ versus the coupling detuning $\Delta_c$ and the free-space link distance $d_{\rm Tx–Rx}$.} \label{LO_free_SNR_Delta_c_distance}
		\vspace{-1em}
\end{figure}

Fig.~\ref{LO_free_SNR_Delta_c_distance} shows the probe laser transmission $P_\text{out}$ as a function of the coupling detuning $\Delta_c$ and the free-space link distance $d_{\rm Tx–Rx}$, along with the SNR performance of LO-free Rydberg atomic receiver, $\mathrm{SNR}_{\rm Ry}$ in~(\ref{SNR_LO_free}).  For each distance $d_{\text{Tx-Rx}}$, we extract a distance-dependent RF Rabi frequency ${\Omega _{{\text{RF}}}}\left( d_{\text{Tx-Rx}} \right)$ and then use QuTip toolkit to generate the EIT-AT spectrum as two Lorentzian peaks (half-width $\Gamma_\text{FWHM} /2 $) centered at detunings $\Delta_c = \pm \frac{1}{2}\Omega_{{\text{RF}}} \left( {{d_{{\text{Tx-Rx}}}}} \right)$. At short range ($d_\text{Tx-Rx} \lesssim 100~\text{m}$) and large $\mathcal{R}$, the AT splitting interval can be well resolved. As $d_{\text{Tx-Rx}}$ increases, $\Omega_\text{RF}$ approaches $\Gamma_\text{FWHM}$, causing the two peaks to merge into a single broadened feature and a rapid drop in ${\text{SNR}}_{\text{Ry}}$. Beyond the critical threshold $R=1$, the AT splitting becomes unobservable, defining the receiver's practical detection range. 
Crossing the threshold $\mathcal{R} = 1$ erases the splitting entirely, thus marking the receiver's effective detection range.

\begin{figure*}[t]
	\centering
	\includegraphics[width=0.8\textwidth]{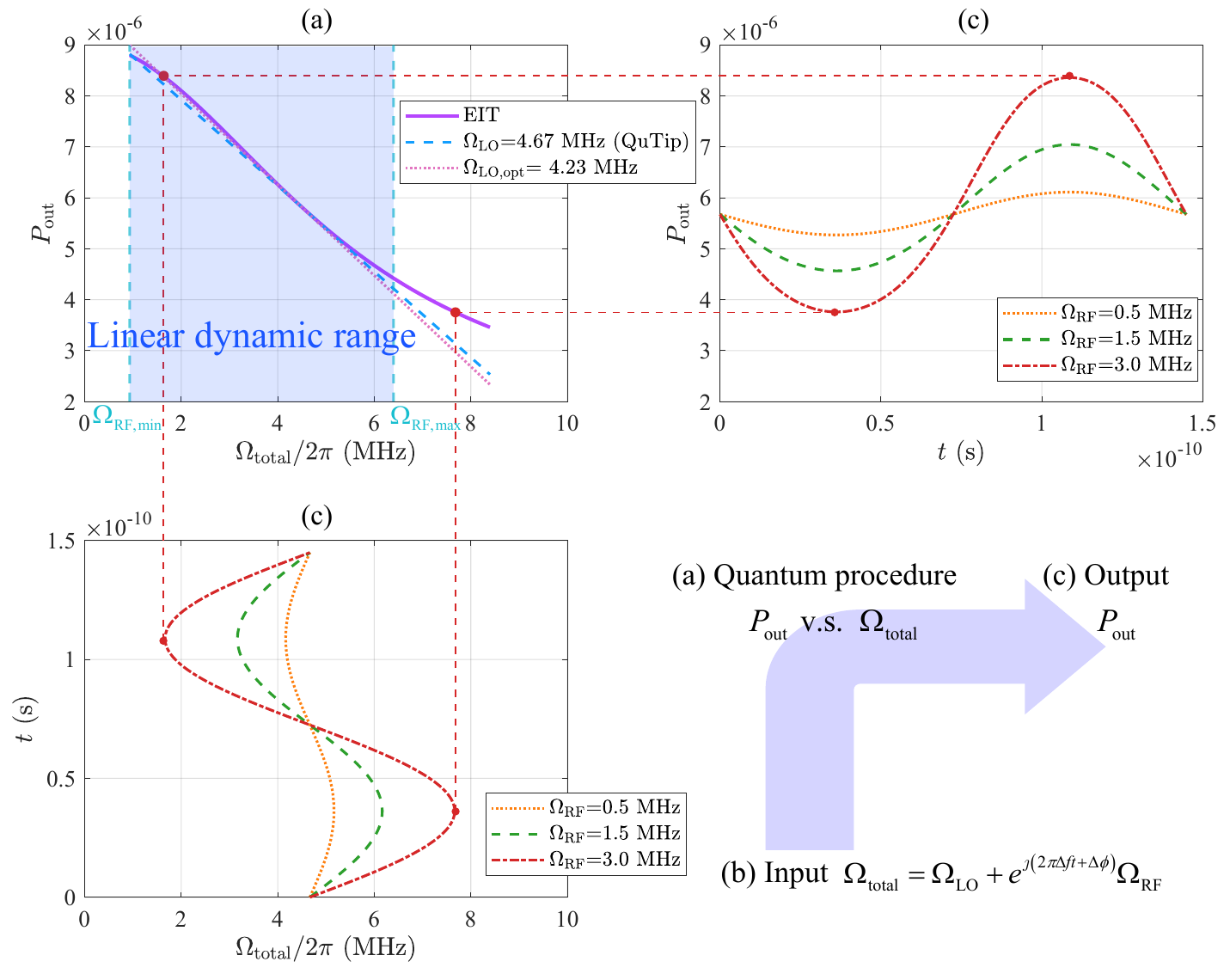}
	\caption{Linear dynamic range of the LO-dressed Rydberg atomic receiver. (a) The probe transmission $P_\text{out}$ versus the Rabi frequency of the superposition field $\Omega_{{\text{total}}}$. (b) The Rabi frequency of the superposition field $\Omega_{{\text{total}}}$ versus time $t$. (c) The probe transmission $P_\text{out}$ versus time $t$.} \label{LO_Dressed_distortion}
		\vspace{-1em}
\end{figure*}

\begin{figure}[t]
	\centering
	\includegraphics[width=0.45\textwidth]{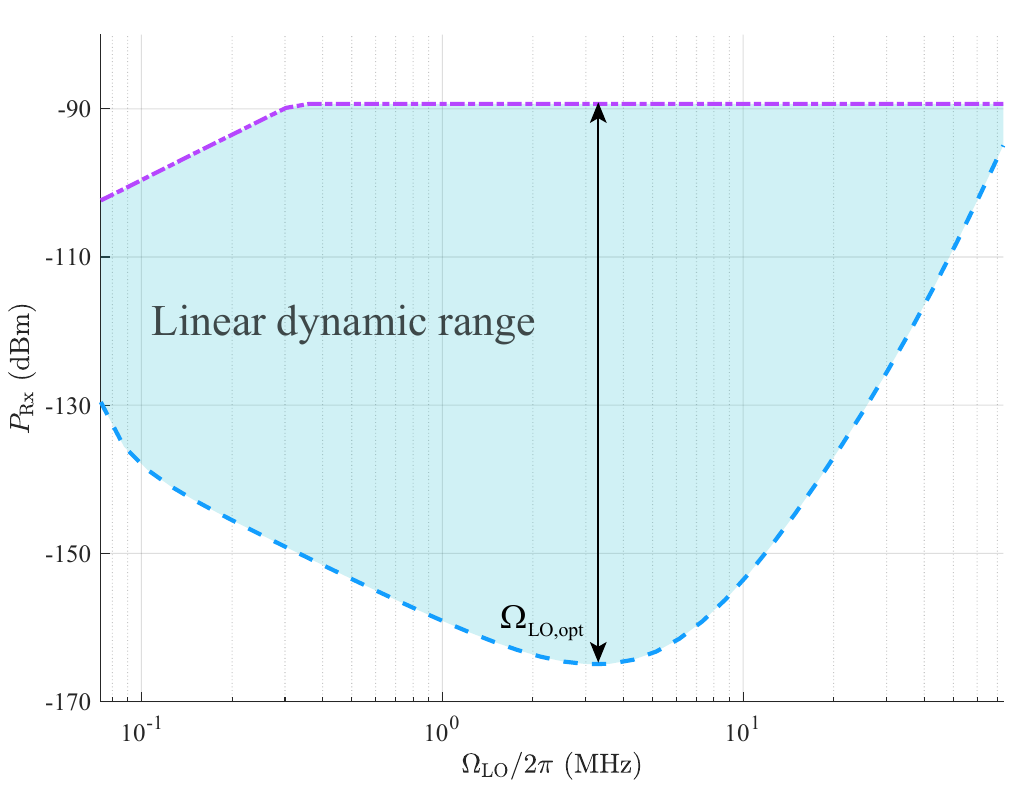}
	\caption{Received power map of the LO-dressed Rydberg atomic receiver.} \label{LO_Dressed_P_Rx_LDR}
	\vspace{-1em}
\end{figure}

\subsubsection{LO-Dressed Rydberg Atomic Receiver}

Given that the atomic response is inherently non-linear, the down-converted photocurrent remains strictly proportional to the signal only inside a finite Rabi frequency range, also referred to as the linear dynamic range. This subsection analyzes this range and links it to the slope ${\kappa _\rho } = {{\partial \Lambda \left( {{\Omega _{{\text{LO}}}},\Gamma } \right)}}/{{\partial {\Omega _{{\text{LO}}}}}}$ that appears in the intrinsic coefficient $\kappa  = \alpha {{\bar P}_0}{\kappa _\rho }$. To be specific, recall that the total Rabi frequency is given by ${\Omega _{{\text{total}}}} = {\Omega _{{\text{LO}}}} + {\Omega _{{\text{RF}}}}{e^{\jmath \theta \left( t \right)}} $ ($\theta \left( t \right) = 2\pi \Delta ft + \Delta \phi$), whose magnitude contains, after a binomial expansion, the undesired high-order harmonics. To make the expression more intuitive, we restructure the probe laser transmission $P_\text{out}$ as
\begin{equation}
{P_{{\text{out}}}} = {P_{{\text{in}}}}{e^{ - \alpha }}{e^{\alpha \Lambda \left( {{\Omega _{{\text{total}}}},\Lambda } \right)}}.
\end{equation}
Expanding ${\Lambda \left( {{\Omega _{{\text{total}}}},\Lambda } \right)}$ on the LO Rabi frequency $\Omega_{{\text{LO}}}$ yields that
\begin{equation}\label{expansion_Lambda_0}
\Lambda \left( {{\Omega _{{\text{total}}}},\Lambda } \right) = {\Lambda _0} + \Lambda _0^\prime \delta \Omega  + \frac{1}{2}\Lambda _0^{\prime \prime }\delta {\Omega ^2} + \frac{1}{6}\Lambda _0^{\prime \prime \prime }\delta {\Omega ^3} + ...,
\end{equation}
where $\delta \Omega  = {\Omega _{{\text{total}}}} - {\Omega _{{\text{LO}}}}$, ${\Lambda _0^\prime} =  - {{2{\Gamma ^2}{\Omega _{{\text{LO}}}}}}/{{{{\left( {{\Gamma ^2} + \Omega _{{\text{LO}}}^2} \right)}^2}}}$, and ${\Lambda_0^{\prime \prime}} =  - {{2{\Gamma ^2}\left( {{\Gamma ^2} - 3\Omega _{{\text{LO}}}^2} \right)}}/{{{{\left( {{\Gamma ^2} + \Omega _{{\text{LO}}}^2} \right)}^3}}}$. Substituting (\ref{expansion_Lambda_0}) into ${P_{{\text{out}}}}$ and retaining terms up to third order yields
\begin{align}
{P_{{\text{out}}}} =& {{\bar P}_0} + \underbrace {\kappa {\Omega _{{\text{RF}}}}\cos \theta }_{{\text{Desired}}} + \underbrace {\frac{1}{4}\alpha {{\bar P}_0}\Lambda _0^{\prime \prime }\delta \Omega _{{\text{RF}}}^2\left( {1 + \cos 2\theta } \right)}_{{\text{Second-order distortion}}} \nonumber\\
  &+ \underbrace {\frac{1}{6}\alpha {{\bar P}_0}\Lambda _0^{\prime \prime }\delta \Omega _{{\text{RF}}}^3{{\cos }^3}\theta }_{{\text{Third-order distortion}}}.
\end{align}
Then, we define $\epsilon$ as the total harmonic distortion (THD)~\cite{THD_definition}. When we compare the amplitudes associated with the fundamental harmonic  $P \left( \Delta f \right) = \left| \kappa \right|  \Omega_{{\text{RF}}}$ and the second-order harmonic $P \left( 2\Delta f \right) = \frac{1}{4}\alpha {{\bar P}_0}\Lambda _0^{\prime \prime }\delta \Omega _{{\text{RF}}}^2$, we have
\begin{equation}
\frac{{P\left( {\Delta f} \right)}}{{P\left( {3\Delta f} \right)}} = \frac{{\kappa {\Omega _{{\text{RF}}}}}}{{\frac{1}{4}\alpha {{\bar P}_0}\Lambda _0^{\prime \prime }\delta \Omega _{{\text{RF}}}^2}} \le \epsilon,
\end{equation}
which results in the second-order response $\Omega _{\text{RF,max}}^{\prime \prime} $
\begin{equation}\label{second_order_RF_Rabi_freq}
{\Omega _{\text{RF,max}}^{\prime \prime} } = \frac{{4 \epsilon \left| {\Lambda _0^\prime } \right|}}{{\left| {\Lambda _0^{\prime \prime }} \right|}}.
\end{equation} 
When the optimal LO Rabi frequency ${\Omega _{{\text{LO,opt}}}} = \frac{{\sqrt 3 }}{3}\Gamma $ is employed, we obtain a maxima of $\kappa_\rho$ given by ${\left( {{\kappa _\rho }} \right)_{\max }} = \frac{{3\sqrt 3 }}{{8\Gamma }}$ (see Appendix~\ref{derivation_P_out}). Under this condition, the second-order distortion vanishes and we need to turn to the third-order distortion, which yields that
\begin{equation}\label{third_order_RF_Rabi_freq}
{\Omega _{\text{RF,max}}^{\prime \prime \prime} } = \sqrt{\frac{{6 \epsilon \left| {\Lambda _0^\prime } \right|}}{{\left| {\Lambda _0^{\prime \prime \prime}} \right|}}}.
\end{equation}

We remark here that the LO-dressed Rydberg atomic receiver is capable of achieving its optimal performance in the linear dynamic range when the LO drives the transition at $\Omega_{{\text{LO,opt}}}$, which simultaneously maximizes the intrinsic coefficient $\kappa$ and eliminates the second-order distortion. The upper bound of the linear dynamic range is determined by $ {\Omega _{\text{RF,max}}^{\prime \prime } } $ (or ${\Omega _{\text{RF,max}}^{\prime \prime \prime} }$), shown in (\ref{second_order_RF_Rabi_freq})-(\ref{third_order_RF_Rabi_freq}). By contrast,  its lower bound corresponds to the intrinsic sensitivity of the LO-dressed Rydberg atomic receiver. Above this limit, the harmonic distortion exceeds the designer-specified tolerance $\epsilon$, although the detector may still operate with post-calibration at the expense of linearity.


Then, simulation results in Fig.~\ref{LO_Dressed_distortion} and Fig.~\ref{LO_Dressed_P_Rx_LDR} are presented to confirm our theoretical analysis. Fig.~\ref{LO_Dressed_distortion} condenses the quantum detection process into three logically linked subfigures. Fig.~\ref{LO_Dressed_distortion}(a) illustrates the probe laser transmission $P_{\text{out}}$ versus the total Rabi frequency $\Omega_{\text{total}}$. When $\Omega_{\text{total}}$  varies versus the time $t$, as seen in Fig.~\ref{LO_Dressed_distortion}(c), the temporal relationship between $P_{\text{out}}$ and $\Omega_{\text{total}}$ is depicted in Fig.~\ref{LO_Dressed_distortion}(b). Specifically, the purple EIT trace in Fig.~\ref{LO_Dressed_distortion}(a) is obtained with QuTiP by solving the Lindblad master equation for the full four-level Hamiltonian and sweeping $\Omega_{{\text{total}}}$. The vertical pink dotted line marks the analytic optimal LO Rabi frequency $\Omega_{{\text{LO,opt}}} = \frac{\sqrt{3}}{3}\Gamma$ and the dashed blued line is the numerically  numerically-optimized $\Omega_{{\text{LO,opt}}}$ obtained in QuTip by maximizing $\left| {\partial {P_{{\text{out}}}}/\partial {\Omega _{{\text{RF}}}}} \right|$ at ${{\Omega _{{\text{RF}}}}} = 0$. Furthermore, the shaded band delimited by $\Omega_{\text{RF},\min}$ and $\Omega_{\text{RF},\max}$ marks the linear dynamic range of the LO-dressed Rydberg atomic receiver. Within this interval, the probe laser transmission $P_{\text{out}}$ varies quasi-linearly with the total Rabi frequency $\Omega_{{\text{total}}}$; the intrinsic coefficient $\kappa$  therefore attains its maximum magnitude and the receiver operates at peak sensitivity. Fig.~\ref{LO_Dressed_distortion} reveals a practical design rule: by employing an optimal LO drive as presented in (\ref{Omega_LO_opt}), the intrinsic coefficient $\kappa$ is maximized while the LDR remains widest. Conversely, an excessively large RF Rabi frequency will push the receiver into the non-linear regime where higher-order distortion appears. At $\Omega_\text{LO,opt}$, the receiver can still accommodate RF amplitudes up to $\Omega _{\text{RF,max}}^{\prime \prime \prime} $ shown in (\ref{third_order_RF_Rabi_freq}).

Fig.~\ref{LO_Dressed_P_Rx_LDR} translates the same LDR into the RF-engineering perspective: the upper purple envelope indicates the maximum received power $P_{\text{Rx,max}}$. Signals above this boundary push the atom into higher-order response and distort the detect envelope. The lower blue dashed envelope traces the minimum detectable power set by PSN and the sensitivity floor of the LO-dressed Rydberg atomic receiver. Obviously, a selection of an optimal $\Omega_{{\text{LO,opt}}}$ leads to the maximum width of the linear dynamic range.



\section{Numerical Results}\label{section_numerical_results}

\subsection{Simulation Setup}

This section provides numerical results for characterizing the performance of a wireless system relying on both LO-free and LO-dressed receivers.  In addition to the quantum parameters presented in Sec.~\ref{section_distortion_effect}, some other parameters associated with wireless systems are specified as follows. We assume that the Tx is equipped with a dipole antenna having an antenna gain of $G_{\text{Tx}} = 2.15~\text{dBi}$. The RF signal is transmitted at a power of $P_{\text{Tx}} = 30~\text{dBm}$. To benchmark the performance against that of classical systems, we consider a conventional superheterodyne RF receiver comprising a receive antenna ($G_{\text{Rx}} = 2.15~\text{dBi}$) and an LNA with a gain of $G_\text{LNA} = 20~\text{dB}$. The effective aperture of the receive antenna is determined by $\lambda^2_{\text{RF}}/ (4\pi)$. The noise includes BBR-induced extrinsic noise and the thermal noise, leading to a total noise power expressed as $\sigma _{{\text{Conv}}}^2 = G_{{\text{LNA}}}^2\sigma _{{\text{ex}}}^2 + \sigma _{{\text{TN}}}^2$. Accordingly, the SNR of the conventional RF receiver is given by ${\text{SN}}{{\text{R}}_{{\text{Conv}}}} = {{{G_{{\text{LNA}}}}{P_{{\text{Rx}}}}{{\left| h \right|}^2}}}/{{\sigma _{{\text{Conv}}}^2}}$.


\subsection{SNR Performance}

\begin{figure}[t]
	\centering
	\includegraphics[width=0.483\textwidth]{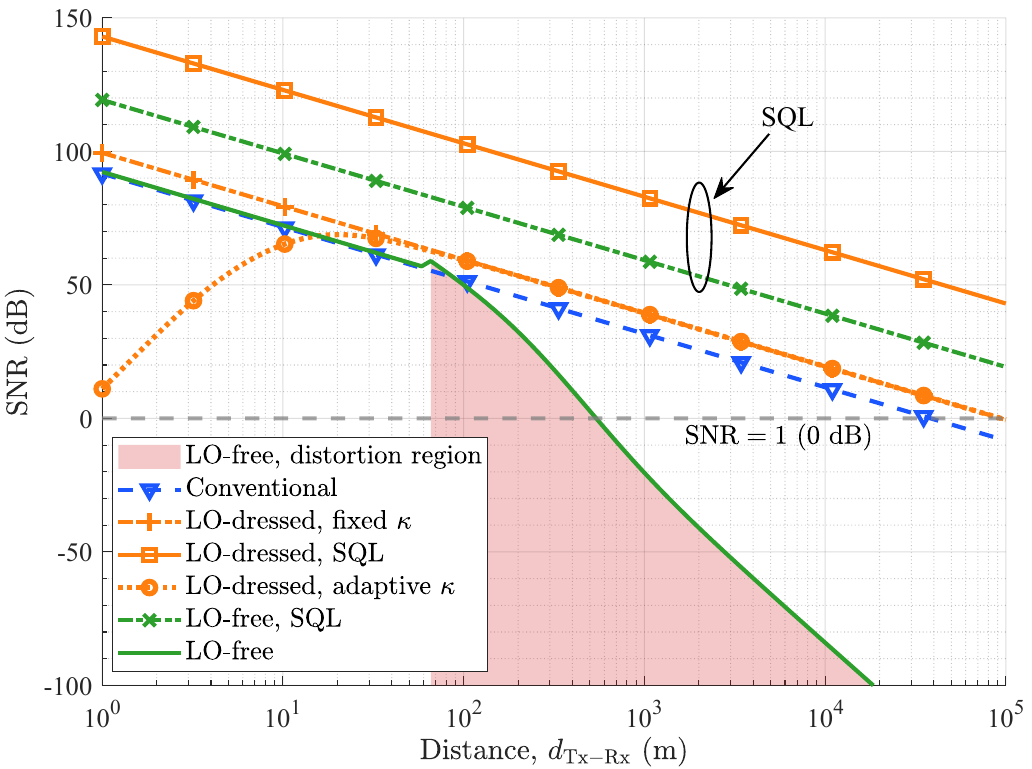}
	\caption{SNR performance versus distance $d_\text{Tx-Rx}$ at different transmit powers $P_\text{Tx}$.} \label{SNR_vs_dist}
	\vspace{-1em}
\end{figure}

Fig.~\ref{SNR_vs_dist} portrays the SNR performance versus distance $d_\text{Tx-Rx}$ of both the LO-free and LO-dressed Rydberg atomic receivers, which are compared to conventional RF receivers. This figure delivers multiple-dimensional perspective, and we now dissect each aspect in turn. Regarding the LO-dressed Rydberg atomic receiver, the `\textit{fixed} $\kappa$' curve is obtained by driving the Rydberg atoms at $\Omega_{{\text{LO,opt}}}$; which assumes that the LO-dressed architecture always operates under ideal coupling and therefore delivers the best‐possible reception. By contrast, the `\textit{adaptive} $\kappa$' curve captures its practical performance. Here the distance $d_\text{Tx-Rx}$ shapes the received power $P_\text{Rx}$ and hence the RF Rabi frequency $\Omega_{{\text{RF}}}$. At small $d_\text{Tx-Rx}$,  $\Omega_{{\text{RF}}}$ may exceed the upper bound of the linear dynamic range determined by $\Omega_ {\text{RF,max}}$ shown in Fig.~\ref{LO_Dressed_distortion}, so the LO-dressed response becomes strongly non-linear and the SNR degrades. As $d_\text{Tx-Rx}$ increases, the received power $P_\text{Rx}$ falls, the receiver re-enters its linear regime, and the SNR rises to a peak. Beyond that point, path-loss dominates and the SNR declines gradually.

We then examine the SNR performance of the LO-free Rydberg atomic receiver. In sharp contrast to the LO-dressed one, it initially delivers a high SNR, but once $d_\text{Tx-Rx}$ grows beyond a certain value that corresponds to a particular RF Rabi frequency $\Omega_{{\text{RF}}}$, the system enters the distortion range. Although the received SNR remains large, the geometric visibility of the AT splitting deteriorates, making it difficult to extract the RF field amplitude directly from the PD readout. Crucially, the condition $\mathcal{R} < \mathcal{R}_\text{th}$ is a distortion criterion that is independent of the conventional sensitivity floor $E_\text{min}$ defined by $\text{SNR} = 1$. Because the LO-free architecture relies on direct optical readout, its fidelity relies on the interplay between AT-peak spacing and linewidth, whereas its ultimate sensitivity is set by the noise floor. As a result, the receiver may ``hear” the signal loudly in an SNR sense yet still render a severely distorted AT profile that precludes accurate field estimation.

Fig.~\ref{SNR_vs_dist} also allows a fair comparison between the three front-end options under the noise model detailed in Fig.~\ref{noise_model}. As it transpires, the LO-free receiver and a conventional heterodyne chain deliver almost identical SNR performance and are both slightly out-performed by the LO-dressed architecture. This can be traced to two-facet facts. Firstly, the Johnson (thermal) noise sets the absolute noise floor and is identical for all front-ends, thus masking subtler noise contributors such as PQN and PSN. Secondly, in the LO-dressed architecture, the detection sensitivity is proportional to the intrinsic slope $\kappa$. The same thermal floor therefore corresponds to a larger effective field-to-voltage gain, which directly translates into the SNR advantage.
Additionally, the curves `\textit{LO-dressed, SQL}' and `\textit{LO-free, SQL}' plot the SQL case specified by (\ref{SQL}). Under these conditions, the LO-dressed Rydberg atomic receiver achieves an SNR roughly 50 dB higher than that of the conventional RF receiver.  This significant performance gain stems entirely from the suppression of QPN by the steep dispersion slope introduced by the dressing tone. In practice, the actual gain will approach this quantum limit only after the electronic noise temperature is driven far below that of the present bench-top set-up, for example by cryogenic read-out or resonantly enhanced optical detection~\cite{QS-3,QS-24}.

\subsection{Mutual Information}

\begin{figure}[t]
	\centering
	\includegraphics[width=0.49\textwidth]{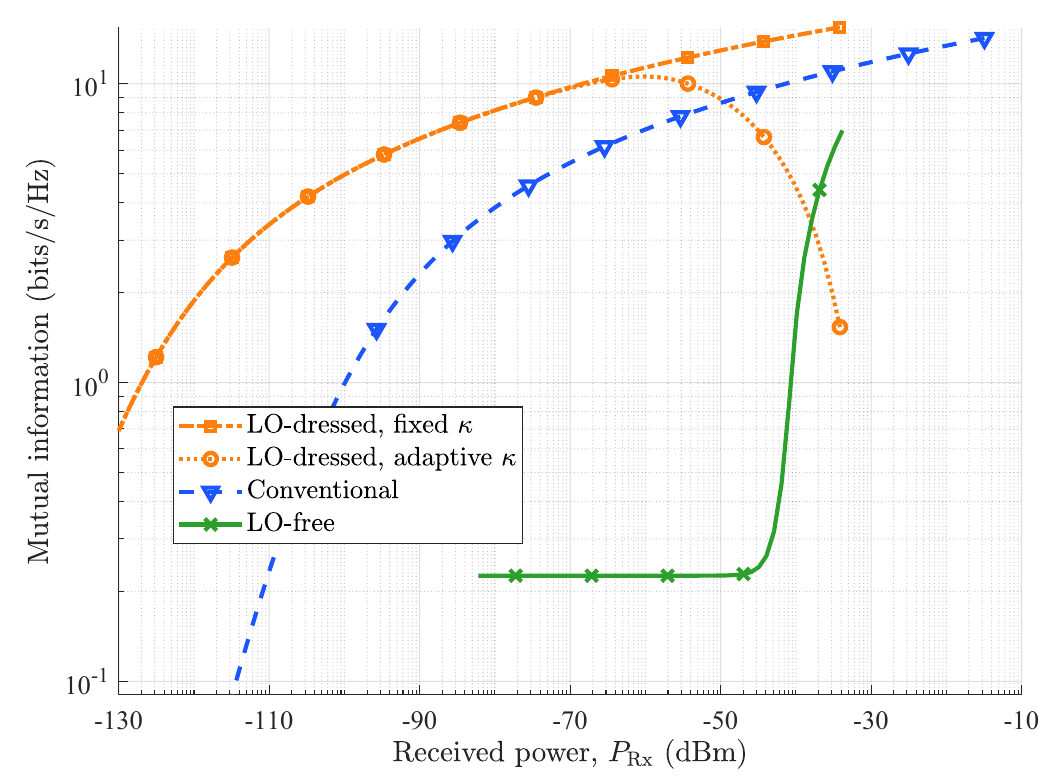}
	\caption{Mutual information versus received power $P_\text{Rx}$.} \label{MI_vs_P_Rx}
\end{figure}

\begin{figure*}[t]
	\centering
	\includegraphics[width=0.81\textwidth]{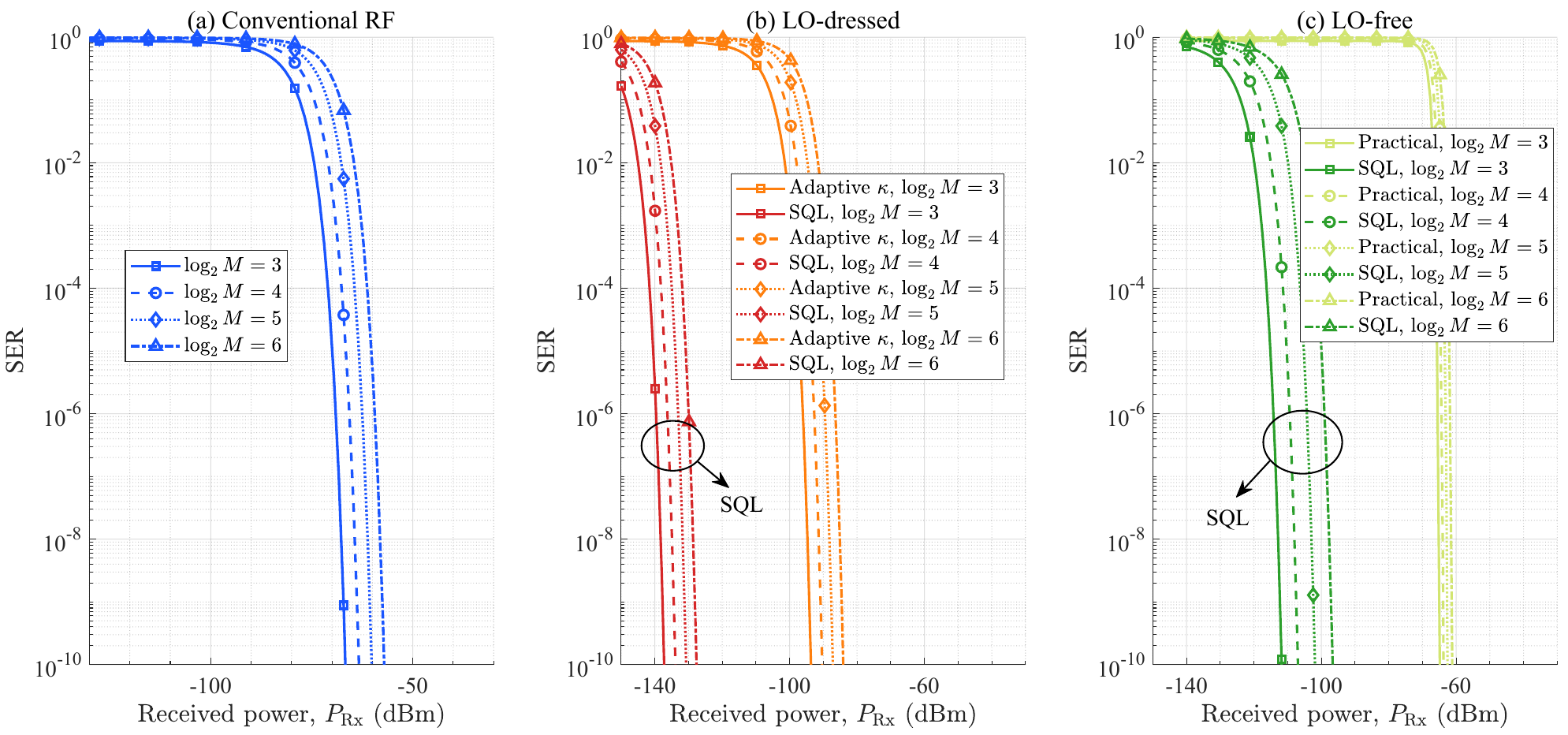}
	\caption{SER performance versus the SNR. (a) Conventional RF receiver. (b) LO-dressed Rydberg atomic receiver. (c) LO-free Rydberg atomic receiver.} \label{SER_vs_SNR}
		\vspace{-1em}
\end{figure*}

Fig.~\ref{MI_vs_P_Rx} illustrates the theoretical mutual information versus the received power $P_\text{Rx}$. According to the signal model presented in (\ref{z_LO_free}), the mutual information $\mathcal{I} \left( z;x\right) $ for the LO-free context is expressed as
\begin{align}\label{mutual_information_LO_free}
	\mathcal{I}\left( {z;x} \right) = \frac{1}{2}\ln \left( {\frac{{\pi (\mathfrak{r} + 1)}}{2}} \right) - \ln {I_0}(\mathfrak{r}) + \mathfrak{r}\frac{{{I_1}(\mathfrak{r})}}{{{I_0}(\mathfrak{r})}},
\end{align}
where ${I_0}\left( \mathfrak{r} \right)$ and ${I_1}\left( \mathfrak{r} \right)$ are the modified Bessel function of the first kind of zero and first orders, respectively; the Rician factor $\mathfrak{r}$ can be approximated by the SNR, i.e., $\mathfrak{r} \approx \text{SNR}_\text{Ry}$. For the LO-dressed one, the mutual information $\mathcal{I}\left( {{z^{{\text{LO}}}};x} \right)$ is calculated by based on the signal model in (\ref{z_LO_dress}), yield that
\begin{equation}\label{LO_dressed_mutual_information}
	\mathcal{I}\left( {{z^{{\text{LO}}}};x} \right) = {e^{1/{\text{SN}}{{\text{R}}_{{\text{Ry,LO}}}}}}{E_1}\left( {1/{\text{SN}}{{\text{R}}_{{\text{Ry,LO}}}}} \right)\ln 2,
\end{equation}
where ${E_1}\left( x \right) = \int_x^\infty  {\frac{{{e^{ - t}}}}{t}{\text{ d}}t} $ denotes the exponential integral function~\cite{math_table}. Their derivations are straightforward, and we omit here for brevity. 
More specifically, given the distortion present in both LO-free and LO-dressed scenarios, their quantum parameters are obtained via QuTiP simulations. As expected, an increase in $P_{\text{Rx}}$ leads to a significant enhancement in mutual information. 
More specifically, the curve `\textit{LO-dressed, adaptive} $\kappa$' is closely aligned with the curve `\textit{LO-dressed, fixed} $\kappa$' at low received power levels. However, as $P_\text{Rx}$ increases, the curve `\textit{LO-dressed, adaptive} $\kappa$' precipitously declines beyond a certain threshold, entering its distortion region. 
By contrast, the LO-free system initially operates within the distortion region at a low received power level, but exhibits substantial capacity improvement beyond this region. These behaviors stem from the distinct atomic responses of LO-free and LO-dressed systems when exposed to varying RF fields. 
Different transmit power levels yield varying RF Rabi frequencies, which in turn results in specific operating or distortion regions at an identical SNR.

In Figs.~\ref{SNR_vs_dist} and \ref{MI_vs_P_Rx}, it may seem counterintuitive that the LO-dressed Rydberg atomic receiver can outperform the conventional one, since the atomic front-end also introduces additional impairments such as QPN, PSN, and BBR-related fluctuations. The key point is that the intrinsic gain associated with the LO-dressed optical readout does not improve the SNR in a naive ``signal-only amplification'' sense. In particular, regarding the LO-dressed noise model presented in (\ref{noise_model_LO_dressed}), the extrinsic input-noise term enters as ${\kappa ^2}{D^2}G_{{\text{LNA}}}^2\sigma _{{\text{ex}}}^2$, which has the same $\kappa$-dependence as the useful signal term in (\ref{SNR_LO_dressed}). We can restructure (\ref{SNR_LO_dressed}) as a more intuitive form
\begin{equation}
{\text{SN}}{{\text{R}}_{{\text{Ry,LO}}}} = \frac{{{R_{\text{L}}}\frac{{\wp _{{\text{RF}}}^2}}{{{\hbar ^2}}}{P_{{\text{Rx}}}}{{\left| h \right|}^2}}}{{{G_{{\text{LNA}}}}\sigma _{{\text{ex}}}^2 + \frac{{{G_{{\text{LNA}}}}\sigma _{{\text{QPN}}}^2}}{{{\kappa ^2}}} + \frac{{{G_{{\text{LNA}}}}\sigma _{{\text{PSN}}}^2}}{{{D^2}{\kappa ^2}}} + \frac{{\sigma _{{\text{TN}}}^2}}{{{D^2}{\kappa ^2}{G_{{\text{LNA}}}}}}}}.
\end{equation}
This form makes the mechanism explicit, i.e., while the common external input noise $\sigma^2_{{\text{ex}}}$ is not reduced by $\kappa^2$, the input-referred contributions of QPN, PSN, and TN are suppressed by the intrinsic gain coefficient $\kappa^2$. {\color{black}Therefore, the SNR advantage of the LO-dressed architecture is most pronounced in practical non-BBR-limited regimes. In the ideal limiting case where both the LO-dressed Rydberg atomic receiver and the conventional RF receiver are limited solely by the same external BBR-related field-fluctuation floor, their ultimate SNR ceilings converge to the same limit. Thus, the advantage observed in Figs.~\ref{SNR_vs_dist} and \ref{MI_vs_P_Rx} should be understood as an improvement in near-SQL regimes, rather than as a violation of a common BBR floor.
}

Therefore, the LO-dressed receiver can achieve a higher overall SNR than a conventional receiver, whenever the operating regime is not purely BBR-limited. In this sense, the advantage of the LO-dressed architecture arises not from violating a common BBR limit, but from reducing the cascaded receiver-induced noise through the high intrinsic gain coefficient $\kappa^2$.

\subsection{SER Performance}

For the SER evaluation, we adopt the models developed in Sec.~\ref{section_modeling_from_phy_to_wireless}. Specifically, for the LO-dressed receiver, the continuous-time readout generated by the atomic-optical mixing stage is assumed to undergo ideal coherent demodulation at the beat frequency $\Delta f = f_{{\text{LO}}} - f_{{\text{RF}}}$, followed by low-pass filtering, matched filtering, and symbol-rate sampling under perfect timing and carrier synchronization. With unit-energy pulse shaping and in the absence of inter-symbol interference, the resultant symbol-spaced observation is characterized by the equivalent discrete-time model in (\ref{z_LO_dress}). Therefore, the SER of the LO-dressed receiver is evaluated using the corresponding standard $M$-QAM error expressions based on the effective SNR derived. By contrast, the LO-free receiver does not produce a continuous complex-valued baseband waveform. Instead, it yields a single block-based amplitude observation per measurement window through scan-and-estimate processing. Accordingly, its SER is evaluated using the corresponding $M$-PAM error expressions based on the effective SNR derived from the LO-free discrete observation model.

Fig.~\ref{SER_vs_SNR} illustrates the SER performance versus the received power $P_\text{Rx}$ for different modulation schemes. The LO-dressed and the conventional systems utilize $M$-order quadrature amplitude modulation ($M$-QAM), while the LO-free systems employ $M$-order pulse amplitude modulation ($M$-PAM) due to its fixed amplitude nature. The SER of PAM is calculated according to \cite[Pg.~266, Eq.~(5.246)]{book-QAM-PAM}, and the SER of the QAM can be evaluated based on \cite[Pg.~278, Eq.~(5.279)]{book-QAM-PAM}.  
Firstly, the three subplots clearly illustrate that, regardless of the receiver type, higher modulation orders result in an increased SER. This aligns with conventional expectations. Crucially, under non-SQL conditions, the LO-dressed Rydberg atomic receiver achieves identical SER with reduced received power $P_\text{Rx}$ attributable to its enhanced field-to-voltage gain determined by the intrinsic coefficient $\kappa$. By contrast, conventional RF and LO-free systems exhibit comparable SER performance. When the analysis is repeated at SQL conditions, both LO-dressed and LO-free systems require approximately $50~\text{dB}$ lower received power than practical implementations to attain equivalent SER. These observations align directly with the trends presented in Fig.~\ref{SNR_vs_dist}.
 
At present, the sensitivity advantage of LO-dressed Rydberg atomic receivers (e.g., $\sim10~\text{nV}/\text{cm}/\sqrt{\text{Hz}}$~\cite{QS-34}) over conventional RF counterparts ($\sim1.88~\text{nV}/\text{cm}/\sqrt{\text{Hz}}$~\cite{terry-trans,3GPP_TR38921}) remains modest. Achieving further sensitivity gains necessitates optimization of quantum parameters such as atomic density, dephasing rate, population rate, resonance linewidth and so on~\cite{terry-trans,QS-22}. Meanwhile, the SNR is significantly constrained by the noise floor. Thus, reducing technical noise, via LO phase noise stabilization and mitigation of optical interferometric drift, is essential. Future work should also explore quantum-limited amplification techniques to approach the fundamental sensitivity bounds of Rydberg-atom based detection.


\section{Conclusions and Future Directions }\label{section_conclusion}
An avenue of realizing true quantum-aided wireless systems has been proposed by developing customized models for Rydberg atomic receivers. We investigated the characteristics of both LO-free and LO-dressed Rydberg atomic receivers, focusing on the linear dynamic range and distortion effects. Through extensive numerical simulations, we characterized our models by evaluating their key performance metrics, including their SNR, system capacities, and SER. Accordingly, a few valuable insights were drawn in terms of their deployment strategies in wireless communication systems.

Despite these contributions, the current convenient framework proposed in this paper represents only the beginning of a broader exploration into quantum sensing aided wireless communications. Some unresolved issues remain, warranting further investigations, and we list a few of them below.
\begin{enumerate}[label={\arabic*)}]
\item \textbf{Approaching SQL regime:} Despite not yet exceeding the raw sensitivity or link-level SNR of leading dipole-antenna front-ends, today's Rydberg atomic receivers are steadily advancing toward the SQL regime~\cite{QS-30,QS-34,QS-9-TAP-2025}. Realizing this regime necessitates parallel advancements in atomic-physics control, in vapor-cell-enabled signal enhancement, and in microwave engineering, particularly with respect to extremely-low-noise optical readout mechanisms and impedance-matched system integration.

\item \textbf{Practical deployment challenges:} Before Rydberg-based receivers can move from laboratory prototypes to wireless fields, several generic issues should be addressed. Specifically, (i) environmental sensitivity: sub-Kelvin temperature drifts, stray magnetic fields, and acoustic vibrations can shift or broaden the EIT resonance. (ii) Hardware constraints: current implementations rely on two frequency-locked diode lasers and external PDs. Advances in photonic integration are reducing the optical package to a compact form factor of just a few $\text{cm}^3$ and $< 1~\text{W}$, comparable to that of low-power superheterodyne front-ends. (iii) System-level integration: dynamic-range management and periodic self-calibration must be co-designed with digital baseband processing.

\item \textbf{Wideband and multi-user extensions:} Recent efforts demonstrate that six-wave-mixing schemes can push Rydberg receivers to continuous bandwidths, while improving atomic-level sensitivity~\cite{QS-30}. Scaling such wideband front-ends to multi-user links, however, faces two open challenges: (i) Spectral multiplexing. A single vapor cell can be driven by multiple LO, each tuned to a different sub-band, enabling orthogonal-frequency or frequency-hopping multiple access. (ii) Spatial multiplexing. Arrays of chip-scale vapor cells, combined with digital beamforming, could separate co-channel users analogously to massive-MIMO antennas.

\end{enumerate}
At this stage, overcoming these engineering barriers, rather than fundamental physical limits, represents the main roadmap for advancing Rydberg atomic receivers toward practical, field-deployable wireless systems. Fortunately, the promise of quantum receivers is generating significant interdisciplinary engagement, particularly among physicists and communications engineers. We contend that this convergence will expedite the maturation of SQL-class Rydberg technologies from laboratory platforms to operational systems.

\ifCLASSOPTIONcaptionsoff
  \newpage
\fi

\appendices
\section{Solution to Master Equation in (\ref{master_equation})}\label{appendix_rho}

To obtain an analytical form of $\rho_{21}$, we first determine the commutator ${\left[ {{\mathbf{H}},\bm{\rho }} \right]_{21}}$. For a $4 \times 4$ Hamiltonian matrix, ${\left[ {{\mathbf{H}},\bm{\rho }} \right]_{ij}}$ is given by ${\left[ {{\mathbf{H}},\bm{\rho }} \right]_{ij}} = \sum\nolimits_{k = 1}^4 {{H_{ik}}{\rho _{kj}} - {\rho _{ik}}{H_{kj}}}$, and thus ${\left[ {{\mathbf{H}},\bm{\rho }} \right]_{21}}$ is formulated as
\begin{equation}\label{commutator_21}
	{\left[ {{\mathbf{H}},\bm{\rho }} \right]_{21}}  = \frac{{\hbar {\Omega _p}}}{2}\left( {{\rho _{11}} - {\rho _{22}}} \right) - \hbar {\Delta _p}{\rho _{21}} + \frac{{\hbar {\Omega _c}}}{2}{\rho _{31}}.
\end{equation}
According to (\ref{master_equation}), we have 
\begin{equation}
	{\dot \rho _{21}} =  - \jmath \frac{{{\Omega _p}}}{2}\left( {{\rho _{11}} - {\rho _{22}}} \right) + \jmath {\Delta _p}{\rho _{21}} - \jmath \frac{{{\Omega _c}}}{2}{\rho _{31}} - {\gamma _{21}}{\rho _{21}}.
\end{equation}
Similarly, we can obtain the evolution equations for  $\rho_{31}$ and $\rho_{41}$, which are respectively given by
\begin{subequations}
	\begin{align}
		{\dot \rho _{31}} =&  - \jmath \frac{{{\Omega _c}}}{2}\left( {{\rho _{21}} - {\rho _{32}}} \right) + \jmath \left( {{\Delta _p} + {\Delta _c}} \right){\rho _{31}} \nonumber\\ 
		&- \jmath \frac{{{\Omega _{{\text{RF}}}}}}{2}{\rho _{41}} - {\gamma _{31}}{\rho _{31}}, \label{rho_31} \\
		{\dot \rho _{41}} =&  - \jmath \frac{{{\Omega _{{\text{RF}}}}}}{2}\left( {{\rho _{31}} - {\rho _{42}}} \right) + \jmath \left( {{\Delta _p} + {\Delta _c} + {\Delta _{{\text{RF}}}}} \right){\rho _{41}} \nonumber \\
		  &- {\gamma _{41}}{\rho _{41}}. \label{rho_41}
	\end{align}
\end{subequations}
The steady-state condition pushes  ${\dot \rho _{21}} = 0$, ${\dot \rho _{31}} = 0$, and ${\dot \rho _{41}} = 0$.
Next, we proceed by eliminating $\rho _{31}$ and $\rho _{41}$ for simplification, and the following approximations are made~\cite{QS-16}: i) given that the population probability of the ground state is approximately 1, i.e., $\rho_{11} \approx 1$, the density components $\rho_{22}$, $\rho_{33}$, and $\rho_{44}$ are relatively small and hence can be neglected; ii) the non-diagonal density components, such as the influence of $\rho_{42}$ and $\rho_{32}$ on $\rho_{21}$, are supposed to be negligible due to their marginal impact. Upon letting ${\rho _{32}} \to 0$ and ${\rho _{42}} \to 0$,  (\ref{rho_31}) and (\ref{rho_41}) can be recast to
\begin{subequations}

\begin{align}
	&\left( {{\gamma _{21}} - \jmath {\Delta _p}} \right){\rho _{21}} =  - \jmath \frac{{{\Omega _p}}}{2}\left( {{\rho _{11}} - {\rho _{22}}} \right) - \jmath \frac{{{\Omega _c}}}{2}{\rho _{31}} , \label{rho_21_tmp}\\
	&\left[ {{\gamma _{31}} - \jmath \left( {{\Delta _p} + {\Delta _c}} \right)} \right]{\rho _{31}} =  - \jmath \frac{{{\Omega _c}}}{2}{\rho _{21}} - \jmath \frac{{{\Omega _{{\text{RF}}}}}}{2}{\rho _{41}}, \label{rho_31_tmp}\\
	&\left[ {{\gamma _{41}} - \jmath \left( {{\Delta _p} + {\Delta _c} + {\Delta _{{\text{RF}}}}} \right)} \right]{\rho _{41}} =  - \jmath \frac{{{\Omega _{{\text{RF}}}}}}{2}{\rho _{31}}. \label{rho_41_tmp}
\end{align}
\end{subequations}
Upon inserting (\ref{rho_41_tmp}) into (\ref{rho_31_tmp}) for eliminating $\rho_{41}$, we have
\begin{equation}\label{rho31_and_rho21}
\resizebox{\hsize}{!}{$
	\left[ {{\gamma _{31}} - \jmath \left( {{\Delta _p} + {\Delta _c}} \right) - \frac{{{{\left( {\frac{{{\Omega _{{\text{RF}}}}}}{2}} \right)}^2}}}{{{\gamma _{41}} - \jmath \left( {{\Delta _p} + {\Delta _c} + {\Delta _{{\text{RF}}}}} \right)}}} \right]{\rho _{31}} =  - \jmath \frac{{{\Omega _c}}}{2}{\rho _{21}}. 
	$}
	\end{equation}
Then, by substituting (\ref{rho31_and_rho21}) into (\ref{rho_21_tmp}), we obtain (\ref{solution_rho_21}), shown at the top of the next page.
	\begin{figure*}[!t]
\begin{align}\label{solution_rho_21}
	\left( {{\gamma _{21}} - \jmath {\Delta _p}} \right){\rho _{21}} =  - \jmath \frac{{{\Omega _p}}}{2}\left( {{\rho _{11}} - {\rho _{22}}} \right) - \jmath \frac{{{\Omega _c}}}{2}\left( {\frac{{ - \jmath \frac{{{\Omega _c}}}{2}}}{ {\gamma _{31}} - \jmath \left( {{\Delta _p} + {\Delta _c}} \right) - \frac{{{{\left( {\frac{{{\Omega _{{\text{RF}}}}}}{2}} \right)}^2}}}{{{\gamma _{41}} - \jmath \left( {{\Delta _p} + {\Delta _c} + {\Delta _{{\text{RF}}}}} \right)}} }{\rho _{21}}} \right) .
\end{align}
\hrule
\end{figure*}
Upon letting ${\rho _{11}} \to 1$ and ${\rho _{22}} \to 0$, we thus obtain the steady-state solution in~(\ref{rho_21_LO_free}).

\section{Derivation of Equivalent Effective Aperture $A_{\rm eff}$ in (\ref{A_eff})}\label{appendix_A_eff}

In contrast to conventional dipole antennas, the effective aperture for a Rydberg atomic receiver is not determined by a geometric area, but instead reflects the quantum-mechanical energy exchange efficiency between the atoms and the incident RF field. This is because the maximum coupling strength between the RF field and the Rydberg transition can be characterized by the Rabi frequency $\Omega_{\text{RF}}$.
Before proceeding with the derivation, we first introduce the steady-state excited-state population $\rho_{ee}$, which is given by~\cite[Sec.~4.2]{support_ref_1}
\begin{equation}
{\rho _{ee}} = \frac{{{{\left( {{\Omega _{{\text{RF}}}}/{\Gamma _{{\text{FWHM}}}}} \right)}^2}}}{{1 + 4{{\left( {{\Delta _c}/{\Gamma _{{\text{FWHM}}}}} \right)}^2} + 2{{\left( {{\Omega _{{\text{RF}}}}/{\Gamma _{{\text{FWHM}}}}} \right)}^2}}},
\end{equation} 
and the total scattering rate is 
\begin{equation}
	\bar R = {{\Gamma _{{\text{FWHM}}}}} \rho _{ee}.
\end{equation}
In the near-resonant weak-limit ($\Delta_c\approx 0, \Omega_{{\text{RF}}} \ll {{\Gamma _{{\text{FWHM}}}}} $), one obtains a scaling given by~\cite{support_ref_1,support_ref_2}
\begin{equation}
	\bar R = \frac{{\Omega _{\text{RF} }^2}}{\Gamma_{\text{FWHM}} } = \frac{{\wp_{\text{RF}}^2}E_{\text{RF}}^2}{{{\hbar ^2}\Gamma_{\text{FWHM}} }}.
\end{equation}
Since each absorption–re-emission event removes an amount of energy $\hbar \omega_{{\text{RF}}}$ from the field, the mean power per atom $P_\text{atom}$ is given by
\begin{equation}
	{P_\text{atom}} = \hbar {\omega _{{\text{RF}}}}\bar R = \frac{{ {\wp_{\text{RF}}^2} E_{\text{RF}}^2 {\omega _{{\text{RF}}}}}}{{\hbar \Gamma_{\text{FWHM}} }}.
\end{equation}
For $N_\text{atoms}$ independent atoms, the total absorbed power becomes
\begin{equation}\label{appendix_P_total}
	{P_{{\text{total}}}} = {N_{{\text{atoms}}}}{P_\text{atom}} = \frac{{{N_{{\text{atoms}}}} {\wp_{\text{RF}}^2} E_{\text{RF}}^2 {\omega _{{\text{RF}}}}}}{{\hbar \Gamma_{\text{FWHM}} }}.
\end{equation}

If we define the power flux density by $\frac{{E_{\text{RF}}^2}}{{2{Z_0}}}$, the total absorbed power in (\ref{appendix_P_total}) can be equivalently structured in the form of 
\begin{equation}
	\frac{{E_{\text{RF}}^2}}{{2{Z_0}}}{A_{{\text{eff}}}} = \frac{{{N_{{\text{atoms}}}}  {\wp_{\text{RF}}^2} E_{\text{RF}}^2 {\omega _{{\text{RF}}}}}}{{\hbar \Gamma_{\text{FWHM}} }}.
\end{equation}
We thus obtain the equivalent effective aperture $A_{\text{eff}}$ formulated as
\begin{align}
	{A_{{\text{eff}}}} = \frac{{2{Z_0}{N_{{\text{atoms}}}}{ \wp_{\text{RF}}^2 }{\omega _{{\text{RF}}}}}}{{\hbar \Gamma_{\text{FWHM}}}},
\end{align}
which corresponds to the formulation in (\ref{A_eff}). This outcome shows that $A_{\text{eff}}$ scales with the number of atoms $N_\text{atoms}$, the Rydberg-Rydberg transition dipole moment $\wp_{\text{RF}}$, the RF-field frequency $\omega_{{\text{RF}}}$ and with the EIT linewidth $\Gamma_{\text{FWHM}}$, rather than the physical cross-section of the vapor cell.

\section{Derivation of (\ref{P_out_kappa})}\label{derivation_P_out}
To make the density matrix component $\rho_{21}$ presented in (\ref{rho_21_LO_dress}) more intuitive and to expose its underlying physical essence, we restructure it as
\begin{align}\label{Im_rho_21_LO}
&\Im \left[ {{\rho _{21}}\left( {{\Omega _{{\text{RF}}}}} \right)} \right] \nonumber\\
&= \bar A\left[ {1 - \Lambda \left( {{\Omega _{{\text{LO}}}},\Gamma } \right) - {\kappa _\rho }{\Omega _{{\text{RF}}}}\cos \left( {2\pi \Delta ft + \Delta \phi } \right)} \right],
\end{align}
where the function $\Lambda \left( {a,b} \right) = {{{b^2}}}/\left( {{{b^2} + {a^2}}}\right) $  is a normalized Lorentzian function with $a$ as the variable and 
$b$ representing the HWHM, where $2b$ corresponds to the FWHM, and ${\kappa _\rho } = {{\partial \Lambda \left( {{\Omega _{{\text{LO}}}},\Gamma } \right)}}/{{\partial {\Omega _{{\text{LO}}}}}}$ represents the intrinsic gain coefficient associated with the density matrix component $\rho_{21}$.  As regards to $\bar A = {{{\gamma _2}{\Omega _p}}}/\left( {{\gamma _2^2 + 2\Omega _p^2}}\right) $, it can be interpreted as the amplitude of the three-level EIT spectrum, and $\Gamma  = {\Omega _p}\sqrt {2\left( {\Omega _c^2 + \Omega _p^2} \right)/2\left( {\Omega _p^2 + \gamma _2^2} \right)} $ represents the HWHM of a four-level system, which is related to the HWHM of the three-level EIT spectrum, denoted by ${\Gamma^{(3)} _{{\text{HWHM}}}}$.
When the probe light is in resonance with the atomic transition and the electric field of the RF signal is zero, we may arrive at
\begin{align}
&\Im \left[ {{\rho _{21}}\left( {{\Delta _c}} \right)} \right]{|_{\left( {{\Delta _p},{\Delta _{{\text{LO}}}},{\gamma _3},{\gamma _4},{\Omega _{{\text{RF}}}},{\Omega _{{\text{LO}}}}} \right) = 0}} \nonumber\\
& = \frac{{{\gamma _2}{\Omega _p}}}{{\gamma _2^2 + 2\Omega _p^2}}\frac{{\Delta _c^2}}{{\Delta _c^2 + {{\left[ {\left( {\Omega _c^2 + \Omega _p^2} \right)/\left( {2\sqrt {\gamma _2^2 + 2\Omega _p^2} } \right)} \right]}^2}}} \nonumber\\
& = \bar A - \bar A\Lambda \left( {{\Omega _{\text{c}}},{\Gamma^{(3)} _{{\text{HWHM}}}}} \right),
\end{align}
where the HWHM of the three-level EIT spectrum ${\Gamma^{(3)} _{{\text{HWHM}}}}$ is given by
\begin{equation}
{\Gamma^{(3)} _{{\text{HWHM}}}} = \left( {\Omega _c^2 + \Omega _p^2} \right)/\left( {2\sqrt {\gamma _2^2 + 2\Omega _p^2} } \right),
\end{equation}
and may be readily shown that 
\begin{align}
\Gamma  = \frac{{2\sqrt 2 {\Omega _p}}}{{\sqrt {\Omega _c^2 + \Omega _p^2} }}{\Gamma^{(3)} _{{\text{HWHM}}}}.
\end{align}
It can be observed from (\ref{Im_rho_21_LO}) that the essence of the LO-dressed Rydberg atomic receiver lies in the measurement of the Rydberg spectrum. The intrinsic gain coefficient $\kappa_\rho$ represents the first-order derivative of the Lorentzian spectrum, reaching its maximum when ${{\Omega _{{\text{LO}}}}}$ satisfies
${{{\partial ^2}\Lambda \left( {{\Omega _{{\text{LO}}}},\Gamma } \right)}}/{{\Omega _{{\text{LO}}}^2}} = 0$,
which provides the criterion for determining the optimal operating condition of the LO-dressed model. This thus yields a single, physically meaningful solution given by
\begin{equation}\label{Omega_LO_opt}
{\Omega _{{\text{LO,opt}}}} = \frac{{\sqrt 3 }}{3}\Gamma ,
\end{equation}
which thus leads to a maxima of $\kappa_\rho$ given by ${\left( {{\kappa _\rho }} \right)_{\max }} = \frac{{3\sqrt 3 }}{{8\Gamma }}$.

Subsequently, we arrive at the practical probe laser transmission $P_\text{out}\left( t \right) $ represented by
\begin{align}\label{P_out_appendix}
&{P_{{\text{out}}}}\left( t \right) \nonumber\\
&= {P_{{\text{in}}}}{e^{ - k_pL{C_0}\Im \left( {{\rho _{21}}} \right)}} \nonumber\\
&= {P_{{\text{in}}}}{e^{ - k_pL{C_0}\bar A\left[ {1 - \Lambda \left( {{\Omega _{{\text{LO}}}},\Gamma } \right) - {\kappa _\rho }{\Omega _{{\text{RF}}}}\cos \left( {2\pi \Delta ft + \Delta \phi } \right)} \right]}} \nonumber\\
&= {P_{{\text{in}}}}{e^{ - k_pL{C_0}\bar A}}{e^{k_pL{C_0}\bar A\Lambda \left( {{\Omega _{{\text{LO}}}},\Gamma } \right)}}{e^{k_pL{C_0}\bar A{\kappa _\rho }{\Omega _{{\text{RF}}}}\cos \left( {2\pi \Delta ft + \Delta \phi } \right)}} \nonumber\\
&= {P_{{\text{in}}}}{e^{ - \alpha }}{e^{\alpha \Lambda \left( {{\Omega _{{\text{LO}}}},\Gamma } \right)}}{e^{\alpha {\kappa _\rho }{\Omega _{{\text{RF}}}}\cos \left( {2\pi \Delta ft + \Delta \phi } \right)}},
\end{align}
where $\alpha  = k_pL{C_0}\bar A$ models the absorption coefficient of the probe laser. Given that ${\Omega _{{\text{RF}}}} \ll {\Omega _{{\text{LO}}}}$, it holds true that $\alpha {\kappa _\rho }{\Omega _{{\text{RF}}}}\cos \left( {2\pi \Delta ft + \Delta \phi } \right) \ll 1$, thus yielding 
\begin{align}\label{P_out_appendix_standard}
{P_{{\text{out}}}}\left( t \right) &\approx {P_{{\text{in}}}}{e^{ - \alpha }}{e^{\alpha \Lambda \left( {{\Omega _{{\text{LO}}}},\Gamma } \right)}}\left( {1 + \alpha {\kappa _\rho }{\Omega _{{\text{RF}}}}\cos \left( {2\pi \Delta ft + \Delta \phi } \right)} \right) \nonumber\\
&= {{\bar P}_0} + \alpha {{\bar P}_0}{\kappa _\rho }{\Omega _{{\text{RF}}}}\cos \left( {2\pi \Delta ft + \Delta \phi } \right) \nonumber\\
&= {{\bar P}_0} + \kappa {\Omega _{{\text{RF}}}}\cos \left( {2\pi \Delta ft + \Delta \phi } \right),
\end{align}
where ${\bar P_0} = {P_{{\text{in}}}}{e^{ - \alpha }}{e^{\alpha \Lambda \left( {{\Omega _{{\text{LO}}}},\Gamma } \right)}}$ and $\kappa  = \alpha {{\bar P}_0}{\kappa _\rho }$ characterizes the total intrinsic gain coefficient.
Additionally, taking into account the practical probe laser transmission in (\ref{P_out_appendix_standard}), the optimal condition ${{{\partial ^2}\Lambda \left( {{\Omega _{{\text{LO}}}},\Gamma } \right)}}/{{\Omega _{{\text{LO}}}^2}} = 0$ can be slightly adjusted according to
\begin{equation}
\frac{{\partial \kappa }}{{\partial {\Omega _{{\text{LO}}}}}} = \frac{{\partial \left( {{{\bar P}_0}{\kappa _\rho }} \right)}}{{\partial {\Omega _{{\text{LO}}}}}} = \frac{{{\partial ^2}{{\bar P}_0}}}{{\Omega _{{\text{LO}}}^2}} = 0,
\end{equation}
which leads to a single, physically meaningful solution given by
\begin{equation}
{\Omega _{{\text{LO}}}} = \sqrt {\bar A - 1 + \sqrt {{{\bar A}^2} - 2\bar A + 4} } \frac{{\sqrt 3 }}{3}\Gamma .
\end{equation}

\section{Derivation of SNR Expression in (\ref{SNR_LO_free})}\label{appendix_SNR_LO_free}

This appendix derives the SNR expression of the LO-free Rydberg atomic receiver. The outcome shows that $G\left( \mathcal{R} \right)$ is not an prescribed term but the inevitable consequence of information theory applied to the AT-splitting. Specifically, with the RF field present, the steady-state EIT can be modelled, to great accuracy, as two identical Lorentzians centred at $\Delta_c = \pm \Omega_{\text{RF}}/2$~\cite{QS-27,addref-1955,QS-16-23}
\begin{equation}
\mathcal{T}\left( {{\Delta _c};{\Omega _{{\text{RF}}}}} \right) = 1 - \mathcal{C}\sum\limits_{i =  \pm 1} {\frac{{{{({\Gamma _{{\text{FWHM}}}}/2)}^2}}}{{{{\left( {{\Delta _c} - \tfrac{{{\Omega _{{\text{RF}}}}}}{2}i} \right)}^2} + {{\left( {{\Gamma _{{\text{FWHM}}}}/2} \right)}^2}}}} ,
\end{equation}
where $\mathcal{C} \in \left( 0,1 \right) $. The PD counts at each detuning point follow a Poisson distribution ${n_k}\sim{\text{Pois}}\left[ {{\lambda _k} = {N_{{\text{ph}}}}\mathcal{T}\left( {{\Delta _c};{\Omega _{{\text{RF}}}}} \right)} \right]$, with $N_\text{ph}$ representing the expected photon number within the detection bandwidth. {
Treating $\Omega_{\text{RF}}$ as the parameter to be estimated, the Poisson Fisher information is given by
\begin{align}\label{poisson_FI}
&\mathcal{I}\left( {{\Omega _{{\text{RF}}}}} \right) \nonumber\\
&= \frac{{{N_{{\text{ph}}}}}}{{2{\Gamma _{{\text{FWHM}}}}}}\int_{ - \infty }^{ + \infty } {\frac{{{{\left[ {{\partial _\mathcal{R}}\mathcal{T}\left( {2{\Delta _c}/{\Gamma _{{\text{FWHM}}}};\mathcal{R}} \right)} \right]}^2}}}{{\mathcal{T}\left( {2{\Delta _c}/{\Gamma _{{\text{FWHM}}}};\mathcal{R}} \right)}}d\left( {\frac{{2{\Delta _c}}}{{{\Gamma _{{\text{FWHM}}}}}}} \right)} ,
\end{align}
where $\mathcal{R} \equiv {{\Omega _{{\text{RF}}}}} / {\Gamma _{{\text{FWHM}}}} $, as shown in~(\ref{ratio_R}).
The Poisson Fisher information $\mathcal{I}\left( {{\Omega _{{\text{RF}}}}} \right)$ in (\ref{poisson_FI}) is bounded as
\begin{align}
\underbrace {\frac{{{N_{{\text{ph}}}}}}{{2{\Gamma _{{\text{FWHM}}}}}}\int_{ - \infty }^{ + \infty } {{{\left[ {{\partial _\mathcal{R}}\mathcal{T}\left( {2{\Delta _c}/{\Gamma _{{\text{FWHM}}}};\mathcal{R}} \right)} \right]}^2}d\left( {\frac{{2{\Delta _c}}}{{{\Gamma _{{\text{FWHM}}}}}}} \right)} }_{{\mathcal{I}_0}\left( \mathcal{R} \right)}\nonumber\\ \le \mathcal{I}\left( {{\Omega _{{\text{RF}}}}} \right) \le \frac{1}{{1 - 2\mathcal{C}}}{\mathcal{I}_0}\left( \mathcal{R} \right),
\end{align}
in which ${\mathcal{I}_0}\left( \mathcal{R} \right)$ has the closed form of 
\begin{equation}
{\mathcal{I}_0}\left( \mathcal{R} \right) = \frac{{8{N_{{\text{ph}}}}{\mathcal{C}^2}}}{{\Gamma _{{\text{FWHM}}}^2}}\frac{1}{{1 + \frac{1}{{2{\mathcal{R}^2}}}}}.
\end{equation}
Then, the exact Fisher information can be written as
\begin{equation}
\mathcal{I}\left( {{\Omega _{{\text{RF}}}}} \right) = \frac{{8{N_{{\text{ph}}}}{\mathcal{C}^2}}}{{\Gamma _{{\text{FWHM}}}^2}}\frac{1}{{1 + \frac{1}{{2{\mathcal{R}^2}}}}}\beta \left( \mathcal{C} \right),
\end{equation}
where $\beta \left( \mathcal{C} \right) \in \left[ {1,\frac{1}{{1 - 2\mathcal{C}}}} \right]$ is a contrast-only factor, independent of $\mathcal{R}$. For typical weak-probe condition adopted in our simulations, e.g., $\mathcal{C} \lesssim 0.2$, we have $\beta \left(\mathcal{C} \right) \le 1.67 $ ($\le2.22~\text{dB}$). Crucially, this factor does not change the $\mathcal{R}$-dependence and can be absorbed in the constant.


}

Then, the Cram{\'{e}}r-Rao bound gives
\begin{equation}
\sigma _{{\Omega _{{\text{RF}}}}}^2 \ge {\mathcal{I}^{ - 1}}\left( {{\Omega _{{\text{RF}}}}} \right) = \frac{{\Gamma _{{\text{FWHM}}}^2}}{{8{N_{{\text{ph}}}}{\mathcal{C}^2}}}\left( {1 + \frac{1}{{2{\mathcal{R}^2}}}} \right).
\end{equation}
In the link budget, the relevant quantity is the power stored in the Rabi frequency $\Omega_{{\text{RF}}}$, leading to an SNR-like expression
\begin{equation}
\frac{{\Omega _{{\text{RF}}}^2}}{{\sigma _{{\Omega _{{\text{RF}}}}}^2}} = \frac{{{\mathcal{R}^2}}}{{1 + \frac{1}{{2{\mathcal{R}^2}}}}} \equiv G\left( \mathcal{R} \right),
\end{equation}
which is exactly the information penalty that quantifies the resolvability loss of the AT splitting.

Next, we multiply the classical wireless link term ${{{P_{{\text{Rx}}}}{{\left| h \right|}^2}}}/{{\sigma _{{\text{Ry}}}^2}}$ by the RF-to-atomic conversion gain ${\left( {{{{\wp _{{\text{RF}}}}}}/{{\hbar {\Gamma _{{\text{FWHM}}}}}}} \right)^2}$ and the Fisher-information penalty $G\left( \mathcal{R} \right)$, yielding the SNR expression presented in (\ref{SNR_LO_free}). Note that the factors ${\left( {\frac{{{\wp _{{\text{RF}}}}}}{{\hbar {\Gamma _{{\text{FWHM}}}}}}} \right)^2}G\left( \mathcal{R} \right)$ is purely dimensionless and does not change the physical units or the engineering interpretation of the SNR. It only reshapes the numerical value according to the intrinsic response of the LO-free Rydberg atomic receiver.

\bibliographystyle{IEEEtran}
\bibliography{ref/Ref_Rydberg}

\begin{IEEEbiography}[{\includegraphics[width=1in,height=1.25in,clip,keepaspectratio]{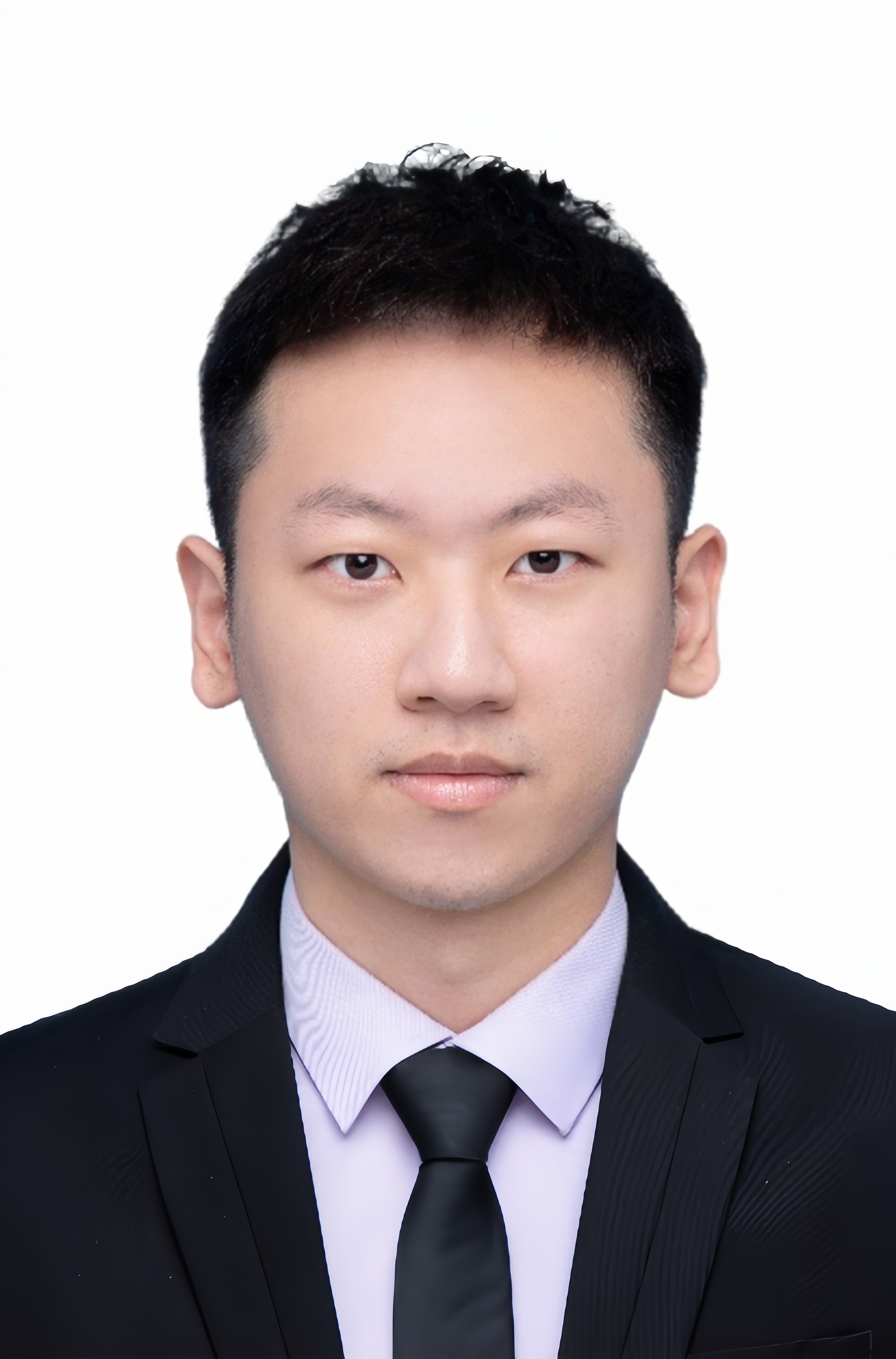}}]{Yuanbin Chen} is a Research Fellow with the School of Electrical and Electronic Engineering, Nanyang Technological University (NTU), Singapore. He received the B.S. degree in Communications Engineering from Beijing Jiaotong University, Beijing, China, in 2019, and the Ph.D. degree in Information and Communication Systems from Beijing University of Posts and Telecommunications (BUPT), Beijing, China, in 2024. He graduated with honors from both institutions.
	
	He was selected in the Stanford/Elsevier Top 2\% Scientists list in 2024 and 2025 and received the Excellent Ph.D. Dissertation Award from the Chinese Institute of Electronics Education Society (2025). He received the 2025 IEEE IWCMC Best Paper Award and the 2023 IEEE GLOBECOM Best Workshop Paper Award. His current research interests include quantum sensing, Rydberg atomic receivers, and 6G advanced MIMO technologies.
\end{IEEEbiography}

\begin{IEEEbiography}[{\includegraphics[width=1in,height=1.25in,clip,keepaspectratio]{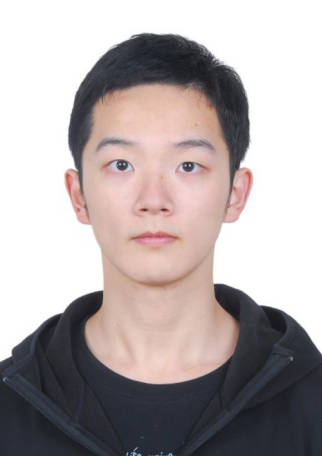}}]{Xufeng Guo} was a visiting Ph.D. student in Nanyang Technological University, Singapore. His research interests include MIMO and quantum sensing.
\end{IEEEbiography}
\vspace{-2.5mm}

\begin{IEEEbiography}[{\includegraphics[width=1in,height=1.25in,clip,keepaspectratio]{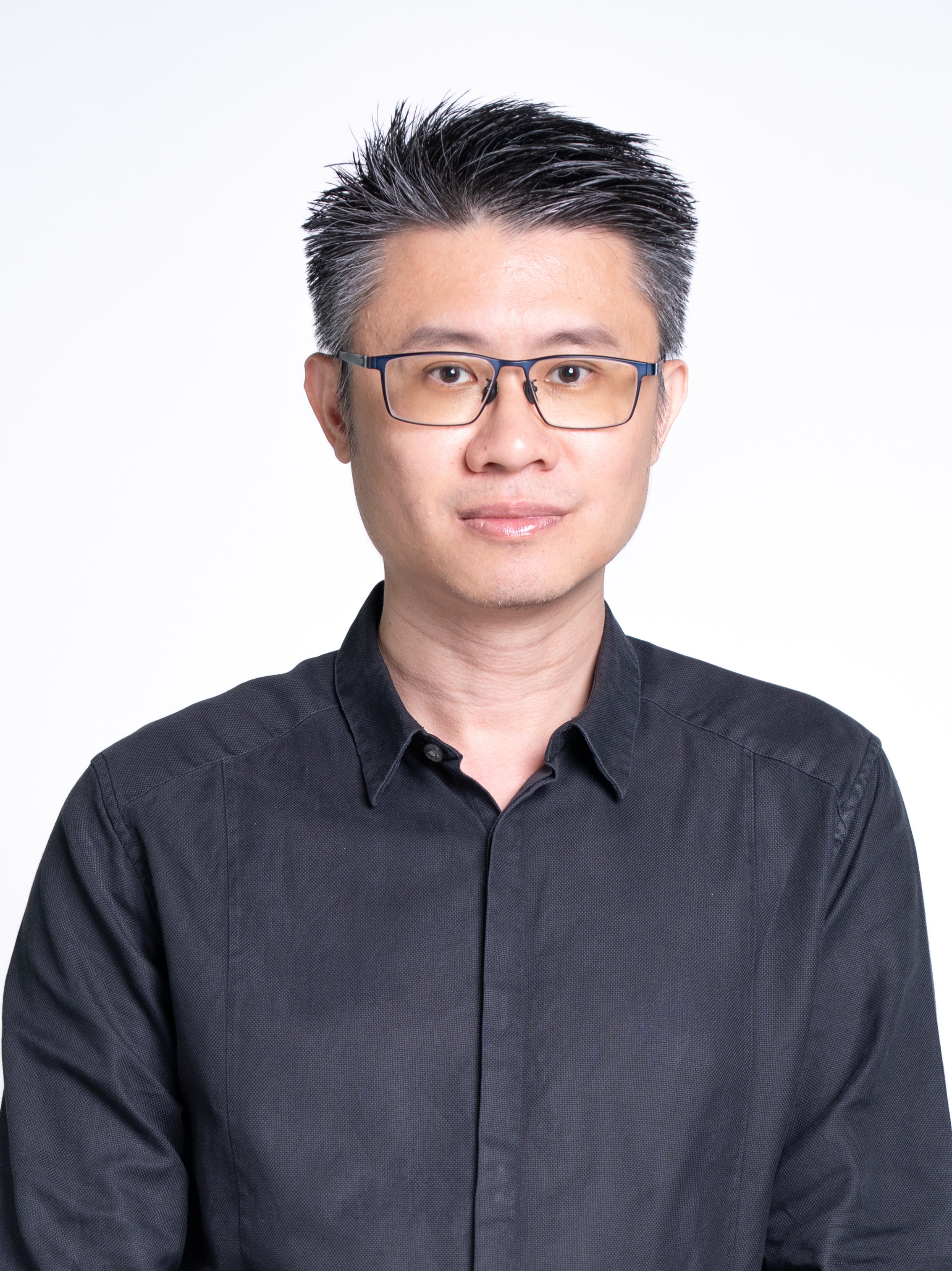}}]{Chau Yuen} received the B.Eng. and Ph.D. degrees from Nanyang Technological University, Singapore, in 2000 and 2004, respectively. He was a Post-Doctoral Fellow with Lucent Technologies Bell Labs, Murray Hill, in 2005. From 2006 to 2010, he was with the Institute for Infocomm Research, Singapore. From 2010 to 2023, he was with the Engineering Product Development Pillar, Singapore University of Technology and Design. Since 2023, he has been with the School of Electrical and Electronic Engineering, Nanyang Technological University, currently he is Provost’s Chair in Wireless Communications, Assistant Dean in Graduate College, and Cluster Director for Sustainable Built Environment at ER@IN. 
	
	Dr. Yuen received IEEE Communications Society Leonard G. Abraham Prize (2024), IEEE Communications Society Best Tutorial Paper Award (2024), IEEE Communications Society Fred W. Ellersick Prize (2023), IEEE Marconi Prize Paper Award in Wireless Communications (2021), IEEE APB Outstanding Paper Award (2023), and EURASIP Best Paper Award for JOURNAL ON WIRELESS COMMUNICATIONS AND NETWORKING (2021).
	
	Dr. Yuen current serves as an Editor for IEEE TRANSACTIONS ON VEHICULAR TECHNOLOGY, IEEE TRANSACTIONS ON NEURAL NETWORKS AND LEARNING SYSTEMS, and IEEE TRANSACTIONS ON NETWORK SCIENCE AND ENGINEERING, where he was awarded as IEEE TNSE Excellent Editor Award 2025, 2024 and 2022, and Top Associate Editor for TVT from 2009 to 2015.
	
	He is listed as Top 2\% Scientists by Stanford University, and also a Highly Cited Researcher by Clarivate Web of Science from 2022.
\end{IEEEbiography}
\vspace{-2.5mm}

\begin{IEEEbiography}[{\includegraphics[width=1in,height=1.25in,clip,keepaspectratio]{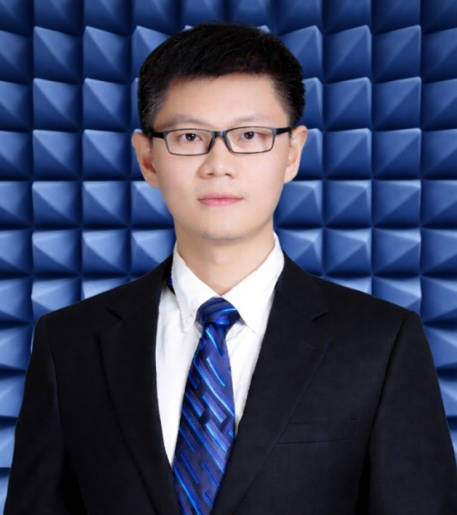}}]{Yufei Zhao} 
	received the B.Eng. degree (Outstanding Graduates) in Electronic Information Engineering from Harbin Institute of Technology, China, in 2014, and the Ph.D. degree in Aeronautical and Astronautical Science and Technology from Tsinghua University, China, in 2020. From 2020 to 2021, he worked as a Senior Engineer at Huawei Technologies Company Ltd., China. From 2021, he worked as a Research Fellow at the School of Electrical and Electronic Engineering, Nanyang Technological University (NTU), Singapore. His current research interests include wireless communication engineering, Rydberg atom technologies, reconfigurable meta-surface, and orbital angular momentum (OAM).
\end{IEEEbiography}

\begin{IEEEbiography}[{\includegraphics[width=1in,height=1.25in,clip,keepaspectratio]{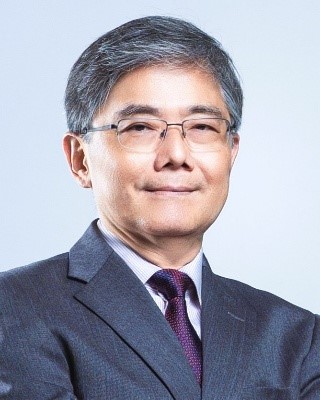}}]{Yong Liang GUAN} obtained his PhD degree from the Imperial College London, UK, and Bachelor of Engineering degree with first class honours from the National University of Singapore.  He is now an Associate Vice President of the Nanyang Technological University (NTU), Singapore, and a Professor of Communication Engineering at the School of Electrical and Electronic Engineering in NTU, where he founded and is leading the Continental-NTU Corporate Research Lab, and also led the successful deployment of the campus-wide NTU-NXP V2X Test Bed in NTU.  His research interests broadly include coding and signal processing for communication systems and data storage systems.  He has published an invited monograph, 2 books and more than 540 journal and conference papers. He has secured over S\$90 million of external research funding.  He has more than 40 filed patents and 13 granted patents (some of which were licensed to NXP, Continental). He is an Editor for the IEEE Transactions on Vehicular Technology and a Distinguished Speaker of the IEEE Vehicular Technology Society.
\end{IEEEbiography}
\vspace{-2.5mm}

\begin{IEEEbiography}[{\includegraphics[width=1in,height=1.25in,clip,keepaspectratio]{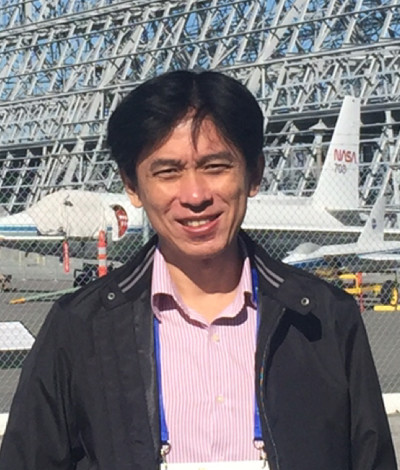}}]{Chong Meng Samson See} received the Diploma in Electronics and Communications Engineering (with Merit) from Singapore Polytechnic, Singapore, in 1988, and the M.Sc. in Digital Communication Systems and the Ph.D. in Electrical Engineering from Loughborough University of Technology, Loughborough, U.K., in 1991 and 1999, respectively.
	
	Since 1992, he has been with DSO National Laboratories, Singapore, where he is currently a Distinguished Member of Technical Staff leading the research and development of advanced array signal processing systems and algorithms. He also holds an adjunct appointment as Principal Research Scientist at Temasek Laboratories @ NTU, Singapore. He holds two issued patents in direction finding.
	His research interests include statistical signal processing, array signal processing, communications, quantum sensing, and the intersection of signal processing and artificial intelligence.
	
Dr. See's contributions to defence research and development have been recognized with the DSO Kinetic Award in 2011 and 2022, the Defence Technology Prize (DTP ) 2014 Team (Engineering) Award, the DTP 2015 Individual (Research and Development) Award, and the DTP 2022 Team (Research) Award.
\end{IEEEbiography}
\vspace{-2.5mm}

\begin{IEEEbiography}[{\includegraphics[width=1in,height=1.25in,clip,keepaspectratio]{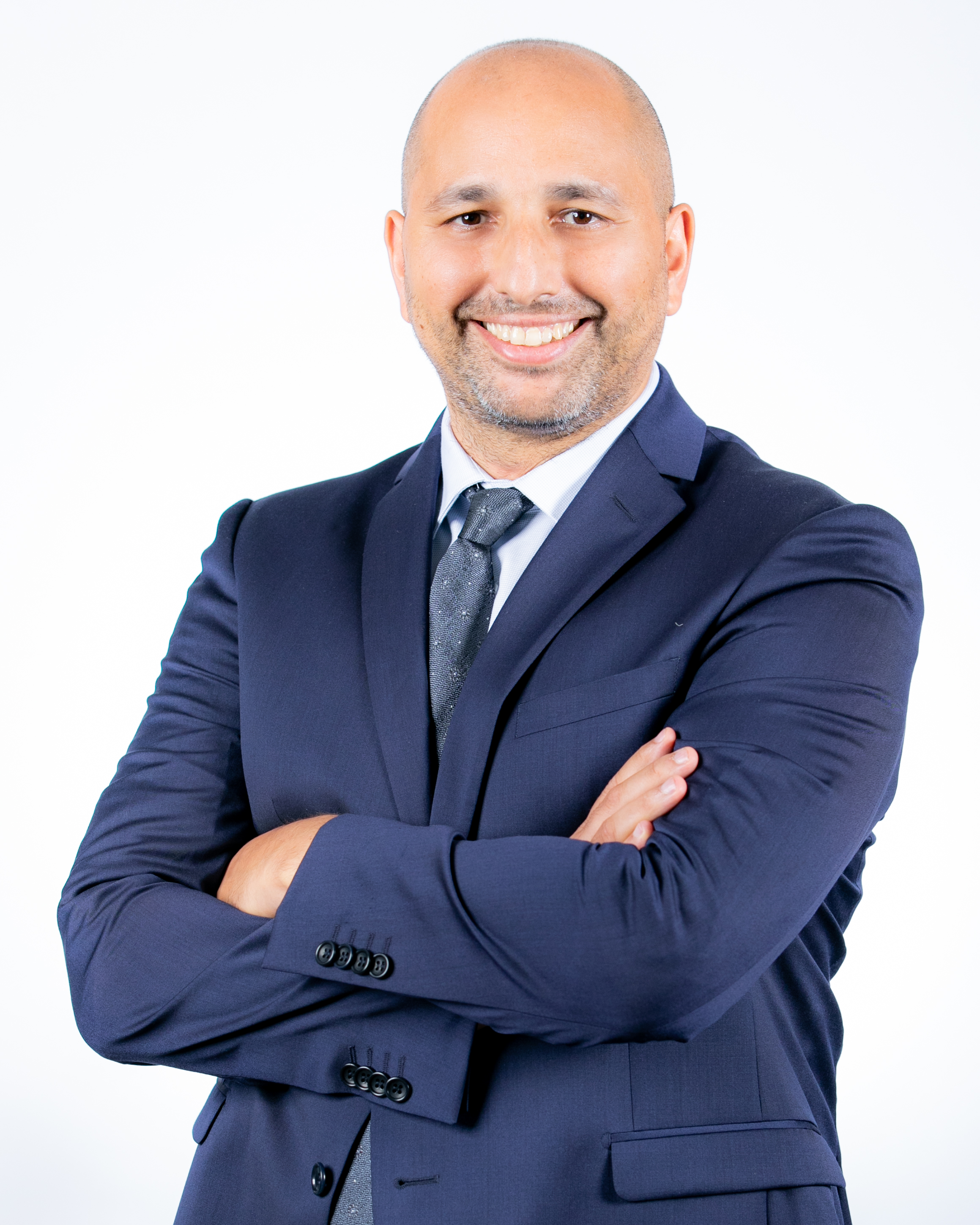}}]{M\'{e}rouane Debbah} is Professor at Khalifa University of Science and Technology in Abu Dhabi and founding Senior Director of KU Digital Future Institute. His research has been lying at the interface of fundamental mathematics, algorithms, statistics, information and communication sciences with a special focus on random matrix theory and learning algorithms. In the Communication field, he has been at the heart of the development of small cells (4G), Massive MIMO (5G) and Large Intelligent Surfaces (6G) technologies. In the AI field, he is known for his work on Large Language Models, distributed AI systems for networks and semantic communications. He received multiple prestigious distinctions, prizes and best paper awards (more than 50 IEEE best paper awards) for his contributions to both fields. He is an IEEE Fellow, a WWRF Fellow, a Eurasip Fellow, an AAIA Fellow, an Institut Louis Bachelier Fellow, an AIIA Fellow  and a Membre émérite SEE. He is actually chair of  the IEEE Large Generative AI Models in Telecom (GenAINet) Emerging Technology Initiative and  a member of the Marconi Prize Selection Advisory Committee.
\end{IEEEbiography}
\vspace{-2.5mm}

\begin{IEEEbiography}[{\includegraphics[width=1in,height=1.25in,clip,keepaspectratio]{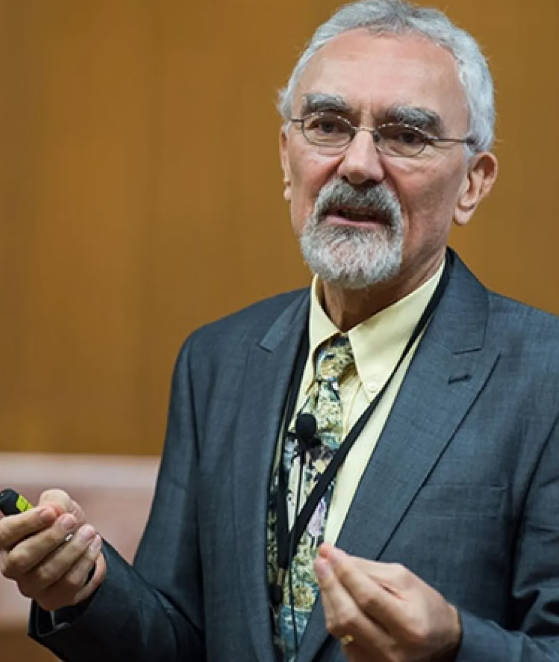}}]{Lajos Hanzo} has co-authored more than 2000 contributions at IEEE Xplore and 19 Wiley-IEEE Press monographs. He is a fellow of the Royal Academy of Engineering, a FIET, a fellow of EURASIP, and a Foreign Member of the Hungarian Academy of Sciences. He was bestowed upon the IEEE Eric Sumner Technical Field Award.
\end{IEEEbiography}

\end{document}